\documentclass[preprintnumbers,amsmath,amssymb,showpacs, twocolumn,superscriptaddress]{revtex4}

\usepackage{graphicx}
\usepackage{subfigure}
\usepackage{dcolumn}
\usepackage{bm}

\begin{document}

\title{Photoproduction of dileptons and photons in p-p collisions at Large Hadron Collider energies}

\author{Zhi-Lei Ma}
\author{Jia-Qing Zhu}
\affiliation {Department of Physics, Yunnan University, Kunming 650091, China}

\affiliation {Key Laboratory of Astroparticle Physics of Yunnan Province, Yunnan University, Kunming 650091, China}

\date{\today}

\begin{abstract}
The production of large $p_{T}$ dileptons and photons originating from photoproduction processes in p-p collisions at
Large Hadron Collider energies is calculated. The comparisons between the exact treatment results and the ones of the
equivalent photon approximation approach are expressed as the $Q^{2}$ (the virtuality of photon) and $p_{T}$ distributions.
The method developed by Martin and Ryskin is used for avoiding double counting when the coherent and incoherent contributions
are considered simultaneously. The numerical results indicate that, the equivalent photon approximation is only effective
in small $Q^{2}$ region and can be used for coherent photoproduction processes with proper choice of $Q^{2}_{\textrm{max}}$ ( the choices
$Q^{2}_{\textrm{max}}\sim \hat{s}$ or $\infty$ will cause obvious errors), but can not be used for incoherent photoproduction processes.
The exact treatment is needed to deal accurately with the photoproduction of large $p_{T}$ dileptons and photons.
\end{abstract}


\maketitle

%**Correspondence author: Jia-Qing Zhu. Email: yndxlyd@163.com;
%Email:867624867@qq.com; 631289401@qq.com;
%Tel:  Zhi-Lei Ma 15198767134; Jia-qing Zhu  15925121909;
\section{INTRODUCTION}

Since its conveniences and simplicity, the equivalent photon approximation (EPA)
, which can be traced to early work by Fermi, Weizs\"{a}cker and Williams (1934),
and Landau and Lifshitz (1934), has been widely used for the approximate calculation
of the various processes in relativistic heavy ion collisions
\cite{Sov.Phys._6_244, Z.Phys._29_315, Z.Phys._88_612, Phys.Rev._45_729, Phys.Rev._51_1037, Phys.Rev._105_1598, Nucl.Phys._23_1295}.
By treating the moving electromagnetic fields of charged particles as a flux of
real photons, many topics are studied such as photoproduction mechanism, particle
and particle pairs production, meson production in electron-nucleon collisions,
two-photon particle production mechanism, the determining of the nuclear parton
distributions, and small x physics \cite{Phys.Rev.C._92_054907, Phys. Rev._104_211, Nucl.Phys.B._900_431, Phys.Rev.C._84_044906, Nucl.Phys.A._865_76, Phys.Rev.C._91_044908, Chin.Phys.Lett._29_081301, Phys.Rep._458_1, Nucl.Phys._53_323, Phys.Rev.D._4_1532,  Phys.Rep._15_181}.
The accuracy of the EPA is denoted by a dynamical cut-off $\Lambda_{\gamma}^{2}$ of
the photon virtuality $Q^{2}$. At $Q^{2}<\Lambda_{\gamma}^{2}$, photo-absorption
cross sections differ slightly from their values on the mass shell and quickly decrease
at $Q^{2}>\Lambda_{\gamma}^{2}$. Thus, the EPA approach is a reasonable approximation
comparing with the exact treatment which returns to the EPA approach when $Q^{2}\rightarrow0$,
and is used precisely for the description of the cross sections only at the kinematics
domain $Q^{2}<\Lambda_{\gamma}^{2}$ \cite{Phys.Rev.D._4_1532, Phys.Lett.B._223_454, Phys.Rev.D._39_2536, Phys.Lett.B._319_339}.
However, the applicability range of EPA and of its accuracy are not always considered in
most works where $Q^{2}_{\textrm{max}}$ is usually set to be $\hat{s}/4$ ($\hat{s}$ is the
squared centre-of-mass (CM) frame energy of the photo-absorption processes) or even infinity,
which will cause a large fictitious contribution from the $Q^{2}>\Lambda_{\gamma}^{2}$
domain \cite{Phys.Rep._15_181}. On the other hand, the EPA approach can not be used for
the study of the incoherent photon emission processes since the parton-quark model is
used which requires $Q^{2}$ should be larger than $\Lambda_{\textrm{QCD}}$, and some
statements in the previous studies \cite{Nucl.Phys.B._900_431, Phys.Rev.C._92_054907, Phys.Rev.C._84_044906, Phys.Rev.C._91_044908, Nucl.Phys.A._865_76,
Chin.Phys.Lett._29_081301} are actually inaccurate.

Hadronic processes for producing large transverse momentum $(p_{T})$ dileptons and photons are very
important in the research of relativistic p-p collisions. Since photons and dileptons do not participate
in the strong interactions directly, the photon or dilepton production can test the predictions of
perturbative quantum chromodynamics (pQCD) calculations, and has long been proposed as ideal probes
of the strong interacting matter (quark-gluon plasma, QGP) properties without the interference of
final-state interactions. In the present work, we extend the photoproduction mechanism which plays
a fundamental role in the \emph{ep} deep inelastic scattering at the Hadron Electron Ring Accelerators \cite{Z.Phys.C._72_637, Phys.Lett.B._377_177, Phys.Rep._345_265, Phys.Rep._332_165} to the production of large $p_{T}$ photons and dileptons in
p-p collisions at Large Hadron Collider (LHC) energies, which has been investigated in most literatures
in the EPA approach. There are several motivations. Although the hard scattering of
initial partons (the annihilation and Compton scattering of partons) is a dominant source of large
$p_{T}$ dileptons and photons in central collisions, the photoproduction processes also playing an
interesting role at LHC energies in which its corrections to the production of dileptons and photons
are non-negligible (especially at the large $p_{T}$ domain) \cite{Phys.Rev.C._84_044906, Phys.Rev.C._91_044908};
The photoproduction processes give the main contribution to the lepton pair production
cross section in the peripheral collisions (especially for the ultra-peripheral collisions);
The EPA is used as an important method in hadronic processes, its validity is not obvious. Therefore,
it is meaningful to study the accuracy of EPA in photoproduction processes and give the accurate
corrections to the production of dileptons and photons.

We present the comparisons between EPA approach and exact treatment in which the photon radiated from
proton or its constituents is off mass shell and no longer transversely polarized, and the minimum of
$p_{T}$ is chosen as $p_{T\ \textrm{min}}=1\ \textrm{GeV}$ for satisfying the requirement of pQCD \cite{Rev.Mod.Phys._59_465, Phys.Rev.C._84_044906}.
There are two types of photoproduction processes: direct photoproduction
processes (dir.pho) and resolved photoproduction processes (res.pho) \cite{Phys.Rev.C._84_044906,
Phys.Rev.C._91_044908}. In the first type, the high-energy photon which are emitted from
proton or the charged parton of the incident proton interacts with the parton of another
incident proton by quark-photon Compton scattering. In the second type, the high-energy
photon, which can be regard as an extend object consisting of quarks and gluons, fluctuates
into a quark-antiquark pair for a short time which then interacts with the parton of
another incident proton by quark-gluon Compton scattering and quark-antiquark annihilation.
Besides, it is necessary to distinguish two kinds of photons emission mechanisms \cite{JPG_24_1657, Nucl.Phys.B._904_386}:
coherent emission in which virtual photons are emitted coherently by the whole proton and the
proton remains intact after the photon radiated; incoherent emission in which virtual photons
are emitted incoherently by the individual constituents (quarks) of proton and the proton will
dissociate or excite after the photon emitted. In most instances, these two kinds of photon
emission processes are performed simultaneously, hence the square of the form factor $F^{2}_{1}(Q^{2})$
is used as coherent probability or weighting factor (WF) for recognizing the coherent part
from the whole interaction. In Ref. \cite{EPJC_74_3040}, Martin and Ryskin have been used this method
to avoid the double counting in the calculation.

The paper is organized as follows. Section. \uppercase\expandafter{\romannumeral2} presents
the formalism of exact treatment for the photoproduction of large $p_{T}$ dileptons and photons
in p-p collisions. Based on the method of Martin and Ryskin, the coherent and incoherent contributions
are considered simultaneously. The EPA approach is also introduced by taking $Q^{2}\rightarrow0$.
In Section. \uppercase\expandafter{\romannumeral3}, the numerical results with the distributions
of $Q^{2}$ and $p_{T}$ at LHC energies are illustrated. Finally, the summary and conclusions are
given in Section. \uppercase\expandafter{\romannumeral4}.
\section{PHOTOPRODUCTION OF LARGE $p_{T}$ DILEPTONS AND PHOTONS}

Since photons and dileptons are the ideal probes in the research of QGP, its production processes
have received many studies in EPA approach. Although the EPA has been widely used as a convenient
method for the approximate calculation of Feynman diagrams for the collision of fast charged particles
\cite{Phys.Rep._15_181}, it's applicability range is often ignored. Thus, the more precise calculations
for the cross sections are needed. We present the exact treatment, which expand the proton or quark tensor
(multiplied by $Q^{-2}$) by using the transverse and longitudinal polarization operators, for the photoproduction
of large $p_{T}$ dileptons and photons in p-p collisions. The formalism is analogous with Refs. \cite{Nucl.Phys.B._621_337,Nucl.Phys.B._904_386}
where the exact treatment for the photoproduction of heavy quarkonia are studied.

\subsection{The $Q^{2}$ distribution of Large $p_{T}$ dilepton production}

The large $p_{T}$ dileptons produced by direct photoproduction processes
can be divided into the coherent direct photoproduction processes (coh.dir)
and incoherent direct photoproduction processes (incoh.dir).
\begin{figure}
\centering
\includegraphics[angle=0,scale=0.30] {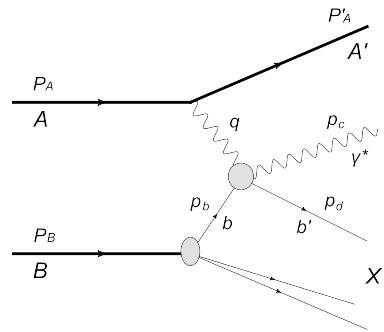}
\vspace*{0.0cm} \caption{The coherent direct photoproduction processes
in which the virtual photon emitted from the whole incident proton A interacts
with parton b of another incident proton B via photon-quark Compton scattering, and
A remains intact after the photon emitted. A$'$ is the scattered proton A,
b$'$ is the scattered parton b, and X is the sum of residue of B after photon
emitted.}
\label{dile.cohdir}
\end{figure}
For the case of coh.dir (Fig. \ref{dile.cohdir}), the invariant cross section of large $p_{T}$ dileptons
with $Q^{2}$ distribution is given by
\begin{eqnarray}\label{d.coh.dir}
&&\frac{d\sigma^{\textrm{coh.dir}}(p+p\rightarrow p+l^{+}l^{-}+X)}{dM^{2}dQ^{2}}\nonumber\\
&=&2\sum_{b}\int dydx_{b}d\hat{t}f_{b/p}(x_{b},\mu_{b}^{2})\frac{d\sigma(p+b\rightarrow p+l^{+}l^{-}+b)}{dM^{2}dQ^{2}dyd\hat{t}},\nonumber\\
\end{eqnarray}
where $x_{b}=p_{b}/P_{B}$ is the parton's momentum fraction, $f_{b/p}(x_{b},\mu_{b}^{2})$
is the parton distribution function of the proton B \cite{Z.Phys.C._53_127}, and the
factorized scale is chosen as $\mu_{b}=\sqrt{4p_{T}^{2}}$ \cite{Phys.Rev.C._84_044906}.
The cross section of the subprocess $p + b\rightarrow p + l^{+}l^{-} +b$ can be written as
\cite{Phys.Rev.D._79_054007, Nucl.Phys.B._904_386}
\begin{eqnarray}\label{sd.coh.dir}
&&\frac{d\sigma(p+b\rightarrow p+l^{+}l^{-}+b)}{dM^{2}dQ^{2}dy}\nonumber\\
&=&\frac{\alpha}{3\pi M^{2}}\sqrt{1-\frac{4m_{l}^{2}}{M^{2}}}(1+\frac{2m_{l}^{2}}
{M^{2}})\frac{d\sigma(p+b\rightarrow p+\gamma^{*}+b)}{dQ^{2}dy}\nonumber\\
&=&\frac{\alpha^{2}}{6\pi^{2}M^{2}}\sqrt{1-\frac{4m_{l}^{2}}{M^{2}}}(1+\frac{2m_{l}^{2}}{M^{2}})
T_{\mu\nu}\frac{y\rho^{\mu\nu}_{\textrm{coh}}}{Q^{2}}\nonumber\\
&&\times\frac{d\textrm{P}\textrm{S}_{2}
(q+p_{b};p_{c},p_{d})}{2yx_{b}s_{NN}},
\end{eqnarray}
where $Q^{2}=-q^{2}$, $y=(q\cdot p_{b})/(P_{A}\cdot p_{b})$, $M$ is the invariant mass of
dileptons, $m_{l}$ is lepton mass, $T_{\mu\nu}$ is the amplitude of reaction $\gamma^{*}+b
\rightarrow\gamma^{*}+b$, $s_{NN}$ is the CM frame energy square of the p-p collision,
$d\textrm{PS}_{2}(q+p_{b};p_{c},p_{d})$ is the Lorentz-invariant phase-space measure
\cite{Nucl.Phys.B._621_337}, the electromagnetic coupling constant is
chosen as $\alpha=1/137$, and
\begin{eqnarray}\label{Rou.coh}
\rho^{\mu\nu}_{\textrm{coh}}&=&(-g^{\mu\nu}+\frac{q^{\mu}q^{\nu}}{q^{2}})H_{2}(Q^{2})\nonumber\\
&&-\frac{(2P_{A}-q)^{\mu}(2P_{A}-q)^{\nu}}{q^{2}}H_{1}(Q^{2}),
\end{eqnarray}
is the proton tensor (multiplied by $Q^{-2}$), $H_{1}(Q^{2})$ and $H_{2}(Q^{2})$ are the elastic form factors of
proton.
\begin{figure}
\centering
\includegraphics[angle=0,scale=0.30] {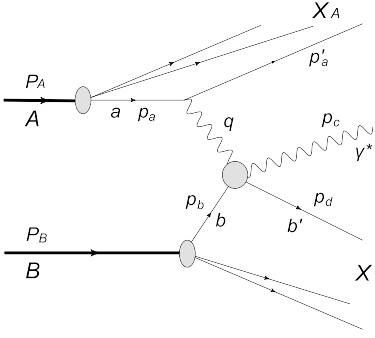}
\vspace*{0.0cm} \caption{The incoherent direct photoproduction processes in which virtual photon
emitted from the parton a of incident proton A interacts with parton b from proton B via the photon-quark
interaction, and A is allowed to break up after photon emitted. a denotes the parton of A, $X_{A}$ is the
residue of A after photon emitted.
}
\label{dile.incohdir}
\end{figure}

Similarly, the invariant cross section of large $p_{T}$ dileptons produced by incoh.dir
(Fig. \ref{dile.incohdir}) with $Q^{2}$ distribution has the form
\begin{eqnarray}\label{d.incoh.dir}
&&\frac{d\sigma^{\textrm{incoh.dir}}(p+p\rightarrow X_{A}+l^{+}l^{-}+X)}{dM^{2}dQ^{2}}\nonumber\\
&=&2\sum_{a,b}\int dydx_{a}dx_{b}d\hat{t}f_{a/p}(x_{a},\mu_{a}^{2})f_{b/p}(x_{b},\mu_{b}^{2})\nonumber\\
&&\times \frac{d\sigma(a+b\rightarrow a+l^{+}l^{-}+b)}{dM^{2}dQ^{2}dyd\hat{t}},
\end{eqnarray}
where $x_{a}=p_{a}/P_{A}$ is parton's momentum fraction, $f_{a/p}(x_{a},\mu_{a}^{2})$ is
the parton distribution function of the proton A, $\mu_{a}=\sqrt{4p_{T}^{2}}$.
And the cross section of the  partonic processes $a + b\rightarrow a + l^{+}l^{-} +b$ reads
\begin{eqnarray}\label{sd.incoh.dir}
&&\frac{d\sigma(a+b\rightarrow a+l^{+}l^{-}+b)}{dM^{2}dQ^{2}dy}\nonumber\\
&=&\frac{\alpha}{3\pi M^{2}}\sqrt{1-\frac{4m_{l}^{2}}{M^{2}}}(1+\frac{2m_{l}^{2}}{M^{2}})\frac{d\sigma(a+b\rightarrow a+\gamma^{*}+b)}{dQ^{2}dy}\nonumber\\
&=&\frac{\alpha^{2}}{6\pi^{2}M^{2}}\sqrt{1-\frac{4m_{l}^{2}}{M^{2}}}(1+\frac{2m_{l}^{2}}{M^{2}})
T_{\mu\nu}\frac{e_{a}^{2}y\rho^{\mu\nu}_{\textrm{incoh}}}{Q^{2}}\nonumber\\
&&\times\frac{d\textrm{P}\textrm{S}_{2}(q+p_{b};p_{c},p_{d})}{2yx_{a}x_{b}s_{NN}},
\end{eqnarray}
where $e_{a}$ is the charge of massless quark a, $y=(q\cdot p_{b})/(p_{a}\cdot p_{b})$ for
the case of incoh.pho, and the massless quark tensor (multiplied by $Q^{-2}$) is
\begin{eqnarray}\label{Rou.incoh.}
\rho^{\mu\nu}_{\textrm{incoh}}&=&(-g^{\mu\nu}+\frac{q^{\mu}q^{\nu}}{q^{2}})L_{2}(Q^{2})\nonumber\\
&&-\frac{(2p_{a}-q)^{\mu}(2p_{a}-q)^{\nu}}{q^{2}}L_{1}(Q^{2}).
\end{eqnarray}

In Martin-Ryskin method \cite{EPJC_74_3040}, the coherent probability (WF) is given by the
square of the form factor $F_{1}^{2}(Q^{2})$. Therefore,
\begin{eqnarray}\label{H}
&&H_{1}(Q^{2})=H_{2}(Q^{2})=F_{1}^{2}(Q^{2}),
\end{eqnarray}
where $F_{1}(Q^{2})$ can be parameterized by the dipole form: $F_{1}(Q^{2})=(1+Q^{2}/0.71\ \textrm{GeV}^{2})^{-2}$.

For the incoherent contribution, the 'remaining' probability has to be considered
for avoiding double counting, and $L_{1}(Q^{2})$, $L_{2}(Q^{2})$
in Eq. (\ref{Rou.incoh.}) have the forms of
\begin{eqnarray}\label{L}
&&L_{1}(Q^{2})=L_{2}(Q^{2})=1-F_{1}^{2}(Q^{2}).
\end{eqnarray}

By using the linear combinations \cite{Phys.Rep._15_181}
\begin{eqnarray}\label{Qmiu}
&&Q^{\mu}=\sqrt{\frac{-q^{2}}{(q\cdot p_{b})^{2}-q^{2}p_{b}^{2}}}(p_{b}-q\frac{q\cdot p_{b}}{q^{2}})^{\mu},\nonumber\\
&&R^{\mu\nu}=-g^{\mu\nu}+\frac{1}{q\cdot p_{b}}(q^{\mu}p_{b}^{\nu}+q^{\nu}p_{b}^{\mu})-\frac{q^{2}}{(q\cdot p_{b})^{2}}p_{b}^{\mu}p_{b}^{\nu},\nonumber\\
\end{eqnarray}
$\rho^{\mu\nu}$ can be written in the form
\begin{eqnarray}\label{Qmiu}
\rho^{\mu\nu}=\rho^{00}Q^{\mu}Q^{\nu}+\rho^{++}R^{\mu\nu}.
\end{eqnarray}
Apparently, $R^{\mu\nu}$ and $Q^{\mu}Q^{\nu}$ are equivalent to transverse and longitudinal
polarization \cite{Nucl.Phys.B._621_337}: $R^{\mu\nu}=\varepsilon_{T}^{\mu\nu}$, $Q^{\mu}Q^{\nu}=-\varepsilon_{L}^{\mu\nu}$.
Thus, the cross section of subprocesses $p+b\rightarrow p+\gamma^{*}+b$ can be expressed as \cite{Nucl.Phys.B._904_386}
\begin{eqnarray}\label{sd.coh.dir.TL}
&&\frac{d^{3}\sigma(p+b\rightarrow p+\gamma^{*}+b)}{dydQ^{2}d\hat{t}}\nonumber\\
&=&\frac{\alpha}{2\pi}[\frac{y\rho^{++}_{\textrm{coh}}}{Q^{2}}\frac{d\sigma_{T}(\gamma^{*}+b\rightarrow\gamma^{*}+b)}{d\hat{t}}\nonumber\\
&&+\frac{y\rho^{00}_{\textrm{coh}}}{Q^{2}}\frac{d\sigma_{L}(\gamma^{*}+b\rightarrow\gamma^{*}+b)}{d\hat{t}}],
\end{eqnarray}
where
\begin{eqnarray}\label{Rouzz00}
&&\rho^{++}_{\textrm{coh}}=F_{1}^{2}(Q^{2})[\frac{1+(1-y)^{2}}{y^{2}}-\frac{2m_{p}^{2}}{Q^{2}}]\nonumber\\
&&\rho^{00}_{\textrm{coh}}=F_{1}^{2}(Q^{2})\frac{4(1-y)}{y^{2}},
\end{eqnarray}
$d\sigma_{T}/d\hat{t}$ and $d\sigma_{L}/d\hat{t}$ represent the transverse and longitudinal
cross sections of subprocesses $\gamma^{*}+b\rightarrow\gamma^{*}+b$ respectively,
\begin{eqnarray}\label{dT.dir.}
&&\frac{d\hat{\sigma}_{T}(\gamma^{*}+b\rightarrow \gamma^{*}+b)}{d\hat{t}}\nonumber\\
&=&\frac{4\pi\alpha^{2}e_{b}^{4}z^{2}}{Q^{4}}[-\frac{\hat{t}}{\hat{s}}-\frac{\hat{s}}{\hat{t}}-M^{2}Q^{2}
(\frac{1}{\hat{s}^{2}}+\frac{1}{\hat{t}^{2}})\nonumber\\
&&+2(Q^{2}-M^{2})\frac{\hat{u}}{\hat{s}\hat{t}}]+\frac{8\pi\alpha^{2}e_{b}^{4}z^{2}}{Q^{4}}
\frac{Q^{2}\hat{u}(\hat{t}-M^2)^2}{\hat{t}^{2}(\hat{s}+Q^{2})^2},\nonumber\\
\end{eqnarray}
and
\begin{eqnarray}\label{dL.dir}
\frac{d\hat{\sigma}_{L}(\gamma^{*}+b\rightarrow \gamma^{*}+b)}{d\hat{t}}
=\frac{8\pi\alpha^{2}e_{b}^{4}z^{2}}{Q^{4}}\frac{Q^{2}\hat{u}(\hat{t}-M^2)^2}
{\hat{t}^{2}(\hat{s}+Q^{2})^2},
\end{eqnarray}
where $z=Q^{2}/(\hat{s}+Q^{2})$, $e_{b}$ is the charge of massless quark b. The Mandelstam variables for subprocesses
$\gamma^{*}+b\rightarrow \gamma^{*}+b$ are defined as
\begin{eqnarray}\label{Mant.dir}
&&\hat{s}=(q+p_{b})^{2}=\frac{M^{2}}{z_{q}}+\frac{p_{T}^{2}}{z_{q}(1-z_{q})},\nonumber\\
&&\hat{t}=(q-p_{c})^{2}=(z_{q}-1)yx_{b}s_{NN},\nonumber\\
&&\hat{u}=(p_{b}-p_{c})^{2}=M^{2}-z_{q}yx_{b}s_{NN},
\end{eqnarray}
where $z_{q}=(p_{c}\cdot P_{B})/(q\cdot P_{B})$ is the inelasticity variable, $p_{T}$ is the transverse momentum in the $\gamma^{*}-b$ CM frame.

In the same way, we can write the cross section of the incoherent subprocesses $a+b\rightarrow a+\gamma^{*}+b$
as
\begin{eqnarray}\label{sd.incoh.dir.TL}
&&\frac{d^{3}\sigma(a+b\rightarrow a+\gamma^{*}+b)}{dydQ^{2}d\hat{t}}\nonumber\\
&=&\frac{\alpha}{2\pi}e_{a}^{2}[\frac{y\rho^{++}_{\textrm{incoh}}}{Q^{2}}\frac{d\sigma_{T}(\gamma^{*}+b\rightarrow\gamma^{*}+b)}{d\hat{t}}\nonumber\\
&&+\frac{y\rho^{00}_{\textrm{incoh}}}{Q^{2}}\frac{d\sigma_{L}(\gamma^{*}+b\rightarrow\gamma^{*}+b)}{d\hat{t}}],
\end{eqnarray}
where
\begin{eqnarray}\label{Rouz0.incoh.}
&&\rho^{++}_{\textrm{incoh}}=(1-F_{1}^{2}(Q^{2}))\frac{1+(1-y)^{2}}{y^{2}}\nonumber\\
&&\rho^{00}_{\textrm{incoh}}=(1-F_{1}^{2}(Q^{2}))\frac{4(1-y)}{y^{2}},
\end{eqnarray}
here we have $\hat{s}=M^{2}/z_{q}+p_{T}^{2}/(z_{q}-z_{q}^{2})$, $\hat{t}=(z_{q}-1)yx_{a}x_{b}s_{NN}$
and $\hat{u}=M^{2}-z_{q}yx_{a}x_{b}s_{NN}$.
\begin{figure}
\centering
\includegraphics[angle=0,scale=0.30] {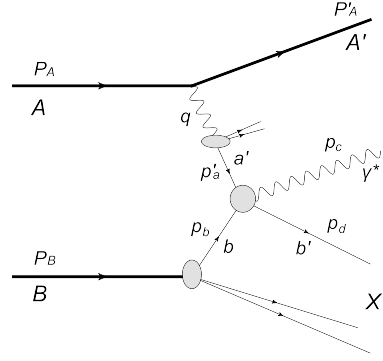}
\vspace*{0.0cm} \caption{The coherent resolved photoproduction processes in which the incident proton A emits a
high energy virtual photon, then the parton $a'$ of the resolved photon interacts with the parton b of
another incident proton B via the interactions of quark-antiquark annihilation and quark-gluon Compton scattering,
and the proton A remains intact after photon emitted.
}
\label{dile.cohres}
\end{figure}

The resolved photoproduction processes are very important in the research of relativistic
heavy ion collisions. The resolved photoproduction processes
can also be divided into two categories: coherent resolved photoproduction processes (coh.res)
and incoherent resolved photoproduction processes (incoh.res). In the case of coh.res (Fig. \ref{dile.cohres}),
the invariant cross section of large $p_{T}$ dileptons with $Q^{2}$ distribution is:
\begin{eqnarray}\label{coh.res.}
&&\frac{d\sigma^{\textrm{coh.res}}(p+p\rightarrow p+l^{+}l^{-}+X)}{dM^{2}dQ^{2}}\nonumber\\
&=&2\sum_{b}\sum_{a'}\int dydx_{b}dz_{a'}d\hat{t}f_{b/p}(x_{b},\mu_{b}^{2})f_{\gamma}(z_{a'},
\mu_{\gamma}^{2})\nonumber\\
&&\times\frac{\alpha}{2\pi}\frac{y\rho^{++}_{\textrm{coh}}}{Q^{2}}\frac
{d\sigma(a'+b\rightarrow \gamma^{*}+b)}{dM^{2}d\hat{t}},
\end{eqnarray}
where $z_{a'}$ denotes the parton's momentum fraction of the resolved photon which are emitted from the proton A,
$f_{\gamma}(z_{a'},\mu_{\gamma}^{2})$ is the parton distribution function of the resolved photon \cite{Phys.Rev.D._60_054019}, $\mu_{\gamma}=\sqrt{4p_{T}^{2}}$.
The cross sections of subprocesses $a'+b\rightarrow\gamma^{*}+b$ are given by
\begin{eqnarray}\label{diledt.cohres}
&&\frac{d\hat{\sigma}}{d\hat{t}}(q\bar{q}\rightarrow\gamma^{*}\gamma)=\frac{2}{3}\frac{\pi\alpha^{2}e_{q}^{4}}{\hat{s}_{\gamma}^{2}}
(\frac{\hat{t}_{\gamma}}{\hat{u}_{\gamma}}+\frac{\hat{u}_{\gamma}}{\hat{t}_{\gamma}}+\frac{2M^{2}\hat{s}
_{\gamma}}{\hat{u}_{\gamma}\hat{t}_{\gamma}})\nonumber\\
&&\frac{d\hat{\sigma}}{d\hat{t}}(q\bar{q}\rightarrow\gamma^{*}g)=\frac{8}{9}\frac{\pi\alpha\alpha_{s}e_{q}^{2}}{\hat{s}_{\gamma}^{2}}
(\frac{\hat{t}_{\gamma}}{\hat{u}_{\gamma}}+\frac{\hat{u}_{\gamma}}{\hat{t}_{\gamma}}+\frac{2M^{2}\hat{s}
_{\gamma}}{\hat{u}_{\gamma}\hat{t}_{\gamma}})\nonumber\\
&&\frac{d\hat{\sigma}}{d\hat{t}}(qg\rightarrow\gamma^{*}q)=\frac{1}{3}\frac{\pi\alpha\alpha_{s}e_{q}^{2}}{\hat{s}_{\gamma}^{2}}
(-\frac{\hat{t}_{\gamma}}{\hat{s}_{\gamma}}-\frac{\hat{s}_{\gamma}}{\hat{t}_{\gamma}}-\frac{2M^{2}\hat{u}
_{\gamma}}{\hat{s}_{\gamma}\hat{t}_{\gamma}}).\nonumber\\
\end{eqnarray}
where $\hat{s}_{\gamma}=M^{2}/z_{q}'+p_{T}^{2}/(z_{q}'-z_{q}'^{2})$, $\hat{t}_{\gamma}=(z_{q}'-1)\hat{s}_{\gamma}$
, and $\hat{u}_{\gamma}=M^{2}-z_{q}'\hat{s}_{\gamma}$ are the Mandelstam variables for the subprocesses of
res.pho, $z_{q}'=(p_{c}\cdot p_{b})/(p_{a'}\cdot p_{b})$ is the inelasticity variable. The strong coupling constant is
taken as the one-loop form \cite{Chin.Phys.Lett._32_121202}
\begin{eqnarray}\label{alphas}
\alpha_{s}=\frac{12\pi}{(33-2n_{f})\ln(\mu^{2}/\Lambda^{2})},
\end{eqnarray}
with $n_{f}=3$ and $\Lambda=0.2\ \textrm{GeV}$.
\begin{figure}
\centering
\includegraphics[angle=0,scale=0.30] {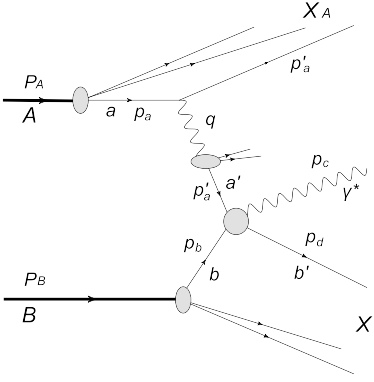}
\vspace*{0.0cm} \caption{The incoherent resolved photoproduction processes in which the parton $a'$ of
the resolved photon radiated by parton a of proton A interacts with the parton b from proton B
via the quark-antiquark annihilation and quark-gluon Compton scattering, and A is allowed to break
up after photon emitted.
}
\label{dile.incohres}
\end{figure}

The invariant cross section for large $p_{T}$ dileptons produced by incoh.res (Fig. \ref{dile.incohres})
with $Q^{2}$ distribution can be presented as
\begin{eqnarray}\label{incoh.res.}
&&\frac{d\sigma^{\textrm{incoh.res}}(p+p\rightarrow X_{A}+l^{+}l^{-}+X)}{dM^{2}dQ^{2}}\nonumber\\
&=&2\sum_{a,b}\sum_{a'}\int dydx_{a}dx_{b}dz_{a'}d\hat{t}f_{a/p}(x_{a},\mu_{a}^{2})f_{b/p}(x_{b},
\mu_{b}^{2})\nonumber\\
&&\times f_{\gamma}(z_{a'},\mu_{\gamma}^{2})e_{a}^{2}\frac{\alpha}{2\pi}\frac{y\rho^{++}_{\textrm{incoh}}}{Q^{2}}\frac
{d\sigma(a'+b\rightarrow \gamma^{*}+b)}{dM^{2}d\hat{t}},
\end{eqnarray}
the cross sections of subprocesses $a'+b\rightarrow\gamma^{*}+b$ are discussed
in Eq. (\ref{diledt.cohres}).
\subsection{The $p_{T}$ distribution of Large $p_{T}$ dileptons production}
It is straightforward to obtain the distribution of $p_{T}$ by accordingly
reordering and redefining the integration variables in Eq. (\ref{d.coh.dir}).
For convenience, the Mandelstam variables in Eq. (\ref{Mant.dir}) can be written
in the form
\begin{eqnarray}\label{Mant.yr}
&\hat{s}&=2\cosh y_{r}M_{T}\sqrt{\cosh^{2}y_{r}M_{T}^{2}-M^{2}}\nonumber\\
&&+2\cosh^{2}y_{r}M_{T}^{2}-M^{2},\nonumber\\
&\hat{t}&=M^{2}-Q^{2}-\sqrt{\hat{s}}M_{T}e^{-y_{r}}+\frac{Q^{2}}{\sqrt{\hat{s}}}M_{T}e^{y_{r}},\nonumber\\
&\hat{u}&=M^{2}-\sqrt{\hat{s}}M_{T}e^{y_{r}}-\frac{Q^{2}}{\sqrt{\hat{s}}}M_{T}e^{y_{r}},
\end{eqnarray}
where $y_{r}=\ln(\cot \theta_{c}/2)$ is the rapidity, $\theta_{c}$ is the CM frame scattering angle, $M_{T}=\sqrt{p_{T}^{2}+M^{2}}$ is the dileptons transverse mass.
By using the Jacobian determinant, the variables $x_{b}$ and $\hat{t}$ can be
transformed into
\begin{eqnarray}\label{Jac.dir.}
d\hat{t}dx_{b}=|\frac{D(x_{b},\hat{t})}{D(p_{T},y_{r})}|dy_{r}dp_{T}^{2},
\end{eqnarray}
where the Jacobian determinant is
\begin{eqnarray}\label{Jacobian.dir.}
&&|\frac{D(x_{b},\hat{t})}{D(p_{T},y_{r})}|\nonumber\\
&=&\frac{\cosh y_{r}}{ys_{NN}\sqrt{\cosh^{2}y_{r}M_{T}^{2}-M^{2}}}[M_{T}(2\cosh y_{r}\nonumber\\
&&+\frac{2\cosh^{2}y_{r}M_{T}^{2}-M^{2}}{M_{T}\sqrt{\cosh^{2}y_{r}M_{T}^{2}-M^{2}}})(M_{T}\sqrt{\hat{s}}-M^{2}e^{y_{r}})\nonumber\\
&&\times(e^{-2y_{r}}+\frac{Q^{2}}{\hat{s}})+\sinh y_{r}(\hat{s}e^{-y_{r}}+\frac{Q^{2}M^{2}e^{y_{r}}}{\hat{s}})\nonumber\\
&&\times(2M_{T}+\frac{2\cosh^{2}y_{r}M_{T}^{2}-M^{2}}{\cosh y_{r}\sqrt{\cosh^{2}y_{r}M_{T}^{2}-M^{2}}})].
\end{eqnarray}

Thus, the invariant cross section of large $p_{T}$ dileptons produced by coh.dir with $p_{T}$
distribution can be expressed as
\begin{eqnarray}\label{ddPT.coh.dir}
&&\frac{d\sigma^{\textrm{coh.dir}}_{y_{rc}}(p+p\rightarrow p+l^{+}l^{-}+X)}{dM^{2}dp_{T}^{2}dy_{r}}\nonumber\\
&=&2\sum_{b}\int dQ^{2}dyf_{b/p}(x_{b},\mu_{b}^{2})\frac{\hat{s}}{ys_{NN}p_{T}}(\sqrt{\hat{s}}+\frac{Q^{2}}{\sqrt{\hat{s}}})\nonumber\\
&&\times\frac{d\sigma(p+b\rightarrow p+l^{+}l^{-}+b)}{dM^{2}dQ^{2}dyd\hat{t}},
\end{eqnarray}
where $y_{rc}$ represents that the cross section is calculated at $y_{r}=0$, the cross section
$d\sigma(p+b\rightarrow p+l^{+}l^{-}+b)/(dM^{2}dQ^{2}dyd\hat{t})$ is discussed in Eq. (\ref{sd.coh.dir})
and Eq. (\ref{sd.coh.dir.TL}).

In the case of incoh.dir, the invariant cross section for large $p_{T}$ dileptons with $p_{T}$ distribution
is given by
\begin{eqnarray}\label{ddPT.incoh.dir}
&&\frac{d\sigma_{y_{rc}}^{\textrm{incoh.dir}}(p+p\rightarrow X_{A}+l^{+}l^{-}+X)}{dM^{2}dp_{T}^{2}dy_{r}}\nonumber\\
&=&2\sum_{a,b}\int dQ^{2}dydx_{a}f_{a/p}(x_{a},\mu_{a}^{2})f_{b/p}(x_{b},\mu_{b}^{2})\nonumber\\
&&\times\frac{\hat{s}}{yx_{a}s_{NN}p_{T}}(\sqrt{\hat{s}}+\frac{Q^{2}}{\sqrt{\hat{s}}})
\frac{d\sigma(a+b\rightarrow a+l^{+}l^{-}+b)}{dM^{2}dQ^{2}dyd\hat{t}},\nonumber\\
\end{eqnarray}
the Mandelstam variables are the same as Eq. (\ref{Mant.yr}), the cross section $d\sigma(a+b\rightarrow a+l^{+}l^{-}+b)/(dM^{2}dQ^{2}dyd\hat{t})$
is discussed in Eq. (\ref{sd.incoh.dir}) and Eq. (\ref{sd.incoh.dir.TL}).

For the case of coh.res, the variables $\hat{t}$ and $z_{a'}$ can be transformed into
\begin{eqnarray}\label{Jac.res.}
d\hat{t}_{\gamma}dz_{a'}=|\frac{D(z_{a'},\hat{t}_{\gamma})}{D(p_{T},y_{r})}|dy_{r}dp_{T}^{2},
\end{eqnarray}
where the Jacobian determinant is
\begin{eqnarray}\label{Jacobian.res.}
&&|\frac{D(z_{a'},\hat{t}_{\gamma})}{D(p_{T},y_{r})}|\nonumber\\
&=&\frac{\cosh y_{r}}{yx_{b}s_{NN}\sqrt{\cosh^{2}y_{r}M_{T}^{2}-M^{2}}}[e^{-2y_{r}}M_{T}(2\cosh y_{r}\nonumber\\
&&+\frac{2\cosh^{2}y_{r}M_{T}^{2}-M^{2}}{M_{T}\sqrt{\cosh^{2}y_{r}M_{T}^{2}-M^{2}}})(M_{T}\sqrt{\hat{s}_{\gamma}}-M^{2}e^{y_{r}})\nonumber\\
&&+\hat{s}_{\gamma}e^{-y_{r}}\sinh y_{r}(\frac{2\cosh^{2}y_{r}M_{T}^{2}-M^{2}}{\cosh y_{r}\sqrt{\cosh^{2}y_{r}M_{T}^{2}-M^{2}}}\nonumber\\
&&+2M_{T})],
\end{eqnarray}
the invariant cross section of large $p_{T}$ dileptons with $p_{T}$ distribution
is given by
\begin{eqnarray}\label{ddPT.coh.res.}
&&\frac{d\sigma_{y_{rc}}^{\textrm{coh.res.}}(p+p\rightarrow p+l^{+}l^{-}+X)}{dM^{2}dp_{T}^{2}dy_{r}}\nonumber\\
&=&2\sum_{b}\sum_{a'}\int dQ^{2}dydx_{b}f_{b/p}(x_{b},\mu_{b}^{2})f_{\gamma}(z_{a'},\mu_{\gamma}^{2})\nonumber\\
&&\times\frac{\hat{s}_{\gamma}^{\frac{3}{2}}}{yx_{b}s_{NN}p_{T}}\frac{\alpha}{2\pi}\frac{y\rho^{++}_{\textrm{coh}}}{Q^{2}}\frac
{d\sigma(a'+b\rightarrow l^{+}l^{-}+b)}{dM^{2}d\hat{t}}.\nonumber\\
\end{eqnarray}

The invariant cross section of large $p_{T}$ dileptons produced
by incoh.res with $p_{T}$ distribution can be written as
\begin{eqnarray}\label{ddPT.incoh.res.}
&&\frac{d\sigma_{y_{rc}}^{\textrm{incoh.res.}}(p+p\rightarrow X_{A}+l^{+}l^{-}+X)}{dM^{2}dp_{T}^{2}dy_{r}}\nonumber\\
&=&2\sum_{a,b}\sum_{a'}\int dQ^{2}dydx_{a}dx_{b}f_{a/p}(x_{a},\mu_{a}^{2})f_{b/p}(x_{b},
\mu_{b}^{2})\nonumber\\
&&\times f_{\gamma}(z_{a'},\mu_{\gamma}^{2})\frac{\hat{s}^{\frac{3}{2}}_{\gamma}}
{yx_{a}x_{b}s_{NN}p_{T}}e_{a}^{2}\frac{\alpha}{2\pi}\frac{y\rho^{++}_{\textrm{incoh}}}{Q^{2}}\nonumber\\
&&\times\frac{d\sigma(a'+b\rightarrow l^{+}l^{-}+b)}{dM^{2}d\hat{t}},
\end{eqnarray}
where the cross sections of subprocesses $a'+b\rightarrow l^{+}l^{-}+b$ are discussed in Eq. (\ref{diledt.cohres}),
and the Mandelstam variables for res.pho are the same as Eq. (\ref{Mant.yr}) but for $Q^{2}=0$.
\subsection{The $Q^{2}$ distribution of Large $p_{T}$ real photon production}
The invariant cross sections of large $p_{T}$ real photons can be derived from
the invariant cross sections of large $p_{T}$ dileptons produced by photoproduction
processes if the invariant mass of dileptons is zero $(M^{2}=0)$. The invariant
cross section of large $p_{T}$ real photons produced by coh.dir with $Q^{2}$
distribution satisfy the following form
\begin{eqnarray}\label{p.coh.dir}
&&\frac{d\sigma^{\textrm{coh.dir}}(p+p\rightarrow p+\gamma+X)}{dQ^{2}}\nonumber\\
&=&2\sum_{b}\int dydx_{b}d\hat{t}f_{b/p}(x_{b},\mu_{b}^{2})\frac{d\sigma(p+b\rightarrow p+\gamma+b)}{dQ^{2}dyd\hat{t}},\nonumber\\
\end{eqnarray}
where the cross section of subprocess $p+b\rightarrow p+\gamma+b$ is
similar to Eq. (\ref{sd.coh.dir.TL}), but the transverse and longitudinal cross
sections of subprocesses $\gamma^{*}+b\rightarrow\gamma+b$ should be presented as
the following forms
\begin{eqnarray}\label{pdT.dir.}
&&\frac{d\hat{\sigma}_{T}}{d\hat{t}}(\gamma^{*}+b\rightarrow \gamma+b)\nonumber\\
&=&\frac{4\pi\alpha^{2}e_{b}^{4}z^{2}}{Q^{4}}[-\frac{\hat{t}}{\hat{s}}-\frac{\hat{s}}{\hat{t}}
+2Q^{2}\frac{\hat{u}}{\hat{s}\hat{t}}]\nonumber\\
&&+\frac{8\pi\alpha^{2}e_{b}^{4}z^{2}}{Q^{2}}\frac{\hat{u}}{(\hat{s}+Q^{2})^2},
\end{eqnarray}
and
\begin{eqnarray}\label{pdL.dir.}
\frac{d\hat{\sigma}_{L}}{d\hat{t}}(\gamma^{*}+b\rightarrow \gamma+b)
=\frac{8\pi\alpha^{2}e_{b}^{4}z^{2}}{Q^{2}}\frac{\hat{u}}
{(\hat{s}+Q^{2})^2},
\end{eqnarray}
where the Mandelstam variables are $\hat{s}=M^{2}/z_{q}+p_{T}^{2}/(z_{q}-z_{q}^{2})$, $\hat{t}=(z_{q}-1)yx_{b}s_{NN}$,
and $\hat{u}=-z_{q}yx_{b}s_{NN}$.

The invariant cross section of large $p_{T}$ real photons produced by incoh.dir
with $Q^{2}$ distribution can be expressed as
\begin{eqnarray}\label{p.incoh.dir}
&&\frac{d\sigma^{\textrm{incoh.dir}}(p+p\rightarrow X_{A}+\gamma+X)}{dQ^{2}}\nonumber\\
&=&2\sum_{a,b}\int dydx_{a}dx_{b}d\hat{t}f_{a/p}(x_{a},\mu_{a}^{2})f_{b/p}(x_{b},\mu_{b}^{2})\nonumber\\
&&\times \frac{d\sigma(a+b\rightarrow a+\gamma+b)}{dQ^{2}dyd\hat{t}},
\end{eqnarray}
here we have $\hat{s}=M^{2}/z_{q}+p_{T}^{2}/(z_{q}-z_{q}^{2})$, $\hat{t}=(z_{q}-1)yx_{a}x_{b}s_{NN}$,
and $\hat{u}=-z_{q}yx_{a}x_{b}s_{NN}$. The partonic cross sections of $a+b\rightarrow a+\gamma+b$
are analogous with Eq. (\ref{sd.incoh.dir.TL}).

In the case of coh.res, the invariant cross section for large $p_{T}$ real photons with
$Q^{2}$ distribution has the form
\begin{eqnarray}\label{p.coh.res.}
&&\frac{d\sigma^{\textrm{coh.res.}}(p+p\rightarrow p+\gamma+X)}{dQ^{2}}\nonumber\\
&=&2\sum_{b}\sum_{a'}\int dydx_{b}dz_{a'}d\hat{t}f_{b/p}(x_{b},\mu_{b}^{2})f_{\gamma}(z_{a'},
\mu_{\gamma}^{2})\nonumber\\
&&\times\frac{\alpha}{2\pi}\frac{y\rho^{++}_{\textrm{coh}}}{Q^{2}}\frac
{d\sigma(a'+b\rightarrow \gamma+b)}{d\hat{t}},
\end{eqnarray}
the cross sections of subprocesses $a'+b\rightarrow\gamma+b$ are given by
\cite{Rev.Mod.Phys._59_465}
\begin{eqnarray}\label{pdT.res.}
&&\frac{d\hat{\sigma}}{d\hat{t}}(q\bar{q}\rightarrow\gamma\gamma)=\frac{2}{3}\frac{\pi\alpha^{2}e_{q}^{4}}{\hat{s}_{\gamma}^{2}}
(\frac{\hat{t}_{\gamma}}{\hat{u}_{\gamma}}+\frac{\hat{u}_{\gamma}}{\hat{t}_{\gamma}}),\nonumber\\
&&\frac{d\hat{\sigma}}{d\hat{t}}(q\bar{q}\rightarrow\gamma g)=\frac{8}{9}\frac{\pi\alpha\alpha_{s}e_{q}^{2}}{\hat{s}_{\gamma}^{2}}
(\frac{\hat{t}_{\gamma}}{\hat{u}_{\gamma}}+\frac{\hat{u}_{\gamma}}{\hat{t}_{\gamma}}),\nonumber\\
&&\frac{d\hat{\sigma}}{d\hat{t}}(qg\rightarrow\gamma q)=\frac{1}{3}\frac{\pi\alpha\alpha_{s}e_{q}^{2}}{\hat{s}_{\gamma}^{2}}
(-\frac{\hat{t}_{\gamma}}{\hat{s}_{\gamma}}-\frac{\hat{s}_{\gamma}}{\hat{t}_{\gamma}}),
\end{eqnarray}
where the Mandelstam variables for the subprocesses of res.pho can be written as $\hat{s}_{\gamma}=M^{2}/z_{q}'+p_{T}^{2}/(z_{q}'-z_{q}'^{2})$,
$\hat{t}_{\gamma}=(z_{q}'-1)\hat{s}_{\gamma}$, and $\hat{u}_{\gamma}=-z_{q}'\hat{s}_{\gamma}$.

For the case of incoh.res, the invariant cross section of large $p_{T}$ real photons
with $Q^{2}$ distribution can be written as
\begin{eqnarray}\label{p.incoh.res.}
&&\frac{d\sigma^{\textrm{incoh.res.}}(p+p\rightarrow X_{A}+\gamma+X)}{dQ^{2}}\nonumber\\
&=&2\sum_{a,b}\sum_{a'}\int dydx_{a}dx_{b}dz_{a'}d\hat{t}f_{a/p}(x_{a},\mu_{a}^{2})f_{b/p}(x_{b},
\mu_{b}^{2})\nonumber\\
&&\times f_{\gamma}(z_{a'},\mu_{\gamma}^{2})e_{a}^{2}\frac{\alpha}{2\pi}\frac{y\rho^{++}_{\textrm{incoh}}}{Q^{2}}\frac
{d\sigma(a'+b\rightarrow \gamma+b)}{d\hat{t}},
\end{eqnarray}
where the cross sections of subprocesses $a'+b\rightarrow\gamma+b$ are
the same as Eq. (\ref{pdT.res.}).
\subsection{The $p_{T}$ distribution of large $p_{T}$ real photon production}
The invariant cross section of large $p_{T}$ real photons produced by coh.dir with $p_{T}$
distribution reads:
\begin{eqnarray}\label{pdPT.coh.dir}
&&E\frac{d\sigma_{y_{rc}}^{\textrm{coh.dir}}(p+p\rightarrow p+\gamma+X)}{d^{3}p}\nonumber\\
&=&\frac{2}{\pi}\sum_{b}\int dQ^{2}dyf_{b/p}(x_{b},\mu_{b}^{2})\frac{\hat{s}}{ys_{NN}p_{T}}(\sqrt{\hat{s}}+\frac{Q^{2}}{\sqrt{\hat{s}}})\nonumber\\
&&\times\frac{d\sigma(p+b\rightarrow p+\gamma+b)}{dQ^{2}dyd\hat{t}},
\end{eqnarray}
the cross sections $d\sigma(p+b\rightarrow p+\gamma+b)/(dQ^{2}dyd\hat{t})$ are discussed in Eq. (\ref{sd.coh.dir.TL})
. And for the case of incoh.dir, the invariant cross section of large $p_{T}$ real photons with $p_{T}$
distribution is:
\begin{eqnarray}\label{pdPT.incoh.dir}
&&E\frac{d\sigma_{y_{rc}}^{\textrm{incoh.dir}}(p+p\rightarrow X_{A}+\gamma+X)}{d^{3}p}\nonumber\\
&=&\frac{2}{\pi}\sum_{a,b}\int dQ^{2}dydx_{a}f_{a/p}(x_{a},\mu_{a}^{2})f_{b/p}(x_{b},\mu_{b}^{2})\nonumber\\
&&\times\frac{\hat{s}}{yx_{a}s_{NN}p_{T}}(\sqrt{\hat{s}}+\frac{Q^{2}}{\sqrt{\hat{s}}})
\frac{d\sigma(a+b\rightarrow a+\gamma+b)}{dQ^{2}dyd\hat{t}},\nonumber\\
\end{eqnarray}
the cross sections $d\sigma(a+b\rightarrow a+\gamma+b)/(dQ^{2}dyd\hat{t})$ are discussed in
Eq. (\ref{sd.incoh.dir.TL}). The Mandelstam variables for dir.pho are the same as Eq. (\ref{Mant.yr}) but for $M^{2}=0$.

The invariant cross section of large $p_{T}$ real photons produced by coh.res with $p_{T}$
distribution can be written as:
\begin{eqnarray}\label{pdPT.coh.res.}
&&E\frac{d\sigma_{y_{rc}}^{\textrm{coh.res.}}(p+p\rightarrow p+\gamma+X)}{d^{3}p}\nonumber\\
&=&\frac{2}{\pi}\sum_{b}\sum_{a'}\int dQ^{2}dydx_{b}f_{b/p}(x_{b},\mu_{b}^{2})f_{\gamma}(z_{a'},
\mu_{\gamma}^{2})\nonumber\\
&&\times\frac{\hat{s}_{\gamma}^{\frac{3}{2}}}{yx_{b}s_{NN}p_{T}}\frac{\alpha}{2\pi}\frac{y\rho^{++}_{\textrm{coh}}}{Q^{2}}\frac
{d\sigma(a'+b\rightarrow\gamma+b)}{d\hat{t}},
\end{eqnarray}
the invariant cross section of large $p_{T}$ real photons produced by incoh.res with $p_{T}$
distribution is given by:
\begin{eqnarray}\label{pdPT.incoh.res.}
&&E\frac{d\sigma_{y_{rc}}^{\textrm{incoh.res.}}(p+p\rightarrow X_{A}+\gamma+X)}{d^{3}p}\nonumber\\
&=&\frac{2}{\pi}\sum_{a,b}\sum_{a'}\int dQ^{2}dydx_{a}dx_{b}f_{a/p}(x_{a},\mu_{a}^{2})f_{b/p}(x_{b},
\mu_{b}^{2})\nonumber\\
&&\times f_{\gamma}(z_{a'},\mu_{\gamma}^{2})\frac{\hat{s}^{\frac{3}{2}}_{\gamma}}
{yx_{a}x_{b}s_{NN}p_{T}}e_{a}^{2}\frac{\alpha}{2\pi}\frac{y\rho^{++}_{\textrm{incoh}}}{Q^{2}}\nonumber\\
&&\times\frac{d\sigma(a'+b\rightarrow \gamma+b)}{d\hat{t}},
\end{eqnarray}
the cross sections of subprocesses $a'+b\rightarrow\gamma+b$ are the same
as Eq. (\ref{pdT.res.}), and the Mandelstam variables for res.pho are the same as Eq. (\ref{Mant.yr}), but for
$Q^{2}=0$ and $M^{2}=0$.
\subsection{The equivalent photons approximation}
The idea of EPA approach is treating the field of a fast charged particle as a flux of photons.
An essential advantage of EPA is that, when using it, it is sufficient to know the photo-absorption
cross section on the mass shell only. Details of its off mass shell behavior are not essential.
Thus, the EPA approach, as a useful technique, has been widely used to obtain the various cross
sections for charged particles in relativistic heavy ion collisions \cite{Phys.Rep._15_181}.
Unfortunately, the accuracy of EPA and its applicability range are often neglected \cite{Phys.Rev.C._84_044906, Phys.Rev.C._91_044908, Nucl.Phys.A._865_76, Chin.Phys.Lett._29_081301}. The choice of $Q^{2}_{\textrm{max}}\sim\hat{s}$ or $\infty$
is used instead of the significant dynamical cut off $\Lambda^{2}_{\gamma}$ which represents
the precision of the EPA approach. However, the exact treatment developed above can be returned
to EPA approach by taking $Q^{2}\rightarrow 0$, and the detailed discussion can be found in Ref. \cite{Phys.Rep._15_181}.
This provides us the powerful comparisons between our results and the ones
in the literatures \cite{Phys.Rev.C._84_044906, Phys.Rev.C._91_044908}. Taking $Q^{2}\rightarrow0$
is corresponding to that the photon is emitted parallelly from proton or quark, and the variable $y$
becomes the usual momentum fraction ($y=q^{+}/P_{A}^{+}$ for coh.pho and $y=q^{+}/p_{a}^{+}$ for incoh.pho) in the
light-front formalism. Since the collinear factorization framework is used for the parton distribution
functions, $x_{a}$ and $x_{b}$ are also equal to $p_{a}^{+}/P_{A}^{+}$ and $p_{b}^{+}/P_{B}^{+}$, respectively.

The cross section of subprocess $p + b\rightarrow p + l^{+}l^{-} +b$ with EPA form reads:
\begin{eqnarray}\label{fgamma.coh.}
\frac{d\sigma^{\textrm{coh.pho}}}{dydQ^{2}d\hat{t}}&=&(\frac{\alpha}{2\pi}\frac{y\rho^{++}_{\textrm{coh}}}{Q^{2}})\frac{d\sigma}{d\hat{t}}\nonumber\\
&=&\frac{\alpha}{2\pi}\frac{F_{1}^{2}(Q^{2})}{Q^{2}}[\frac{1+(1-y)^{2}}{y}-y\frac{2m_{p}^{2}}{Q^{2}}]\frac{d\sigma}{d\hat{t}}\nonumber\\
&=&\frac{df_{\gamma}\shortmid_{\textrm{coh}}(y)}{dQ^{2}}\frac{d\sigma}{d\hat{t}},
\end{eqnarray}
where $f_{\gamma}\shortmid_{\textrm{coh}}(y)$ is the coherent photon flux which is associated
with the whole proton \cite{Nucl.Phys.B._904_386}. It should be noted that, since $\sigma_{T}$
and $\sigma_{L}$ are multiplied by the factor $Q^{-2}$, $\sigma_{L}$ and the terms which are
proportional to $Q^{2}$ in $\sigma_{T}$ can also provide the non-zero contributions when
$Q^{2}\rightarrow 0$, but they are neglected in EPA approach. Actually, the errors from
these omissions are so small, and can not cause any noticeable effects.

Another approximate analytic form of the coherent photon flux is developed
by Drees and Zeppenfeld (DZ) \cite{Phys.Rev.D._39_2536} , which is widely used
in the literatures \cite{Phys.Rev.C._91_044908, Phys.Rev.C._49_1127, Nucl.Phys.B._900_431,
Chin.Phys.Lett._29_081301}. By setting $Q^{2}_{\textrm{max}}\rightarrow\infty$,
and neglecting the $m_{p}^{2}$ term in $\rho^{++}_{\textrm{coh}}$, they obtained
\begin{eqnarray}\label{DZ.fgamma.coh.}
f_{\gamma}\shortmid_{\textrm{coh}}(y)&=&\frac{\alpha}{2\pi}\frac{1+(1-y)^{2}}{y}[\ln A\nonumber\\
&&-\frac{11}{6}+\frac{3}{A}-\frac{3}{2A^{2}}+\frac{1}{3A^{2}}],
\end{eqnarray}
where $A=(1+0.71\ \textrm{GeV}^{2}/Q^{2}_{\textrm{min}})$.

For the case of incoh.pho, the cross section of subprocess $a + b\rightarrow a + l^{+}l^{-} +b$ with
EPA form reads:
\begin{eqnarray}\label{fgamma.incoh.}
\frac{d\sigma^{\textrm{incoh.pho}}}{dydQ^{2}d\hat{t}}&=&(e_{a}^{2}\frac{\alpha}{2\pi}\frac{y\rho^{++}_{\textrm{incoh}}}{Q^{2}})\frac{d\sigma}{d\hat{t}}\nonumber\\
&=&e_{a}^{2}\frac{\alpha}{2\pi}\frac{1-F_{1}^{2}(Q^{2})}{Q^{2}}\frac{1+(1-y)^{2}}{y}\frac{d\sigma}{d\hat{t}}\nonumber\\
&=&\frac{df_{\gamma}\shortmid_{\textrm{incoh}}(y)}{dQ^{2}}\frac{d\sigma}{d\hat{t}},
\end{eqnarray}
where $f_{\gamma}\shortmid_{\textrm{incoh}}(y)$ is the incoherent photon flux.

Another form of Eq. (\ref{fgamma.incoh.}) , which neglects the $F_{1}^{2}(Q^{2})$ term and takes $Q^{2}_{\textrm{min}}=
1\ \textrm{GeV}^{2}$, is
\begin{eqnarray}\label{fgamma.incoh.noWF}
f_{\gamma}\shortmid_{\textrm{incoh}}(y)&=&e_{a}^{2}\frac{\alpha}{2\pi}\frac{1+(1-y)^{2}}{y}\ln\frac{Q^{2}_{\textrm{max}}}{Q^{2}_{\textrm{min}}}.
\end{eqnarray}
\section{NUMERICAL RESULTS}
\begin{figure*}
\begin{tabular}{cc}
\begin{minipage}[t]{1.3in}
\includegraphics[height=8.8cm,width=10.3cm]{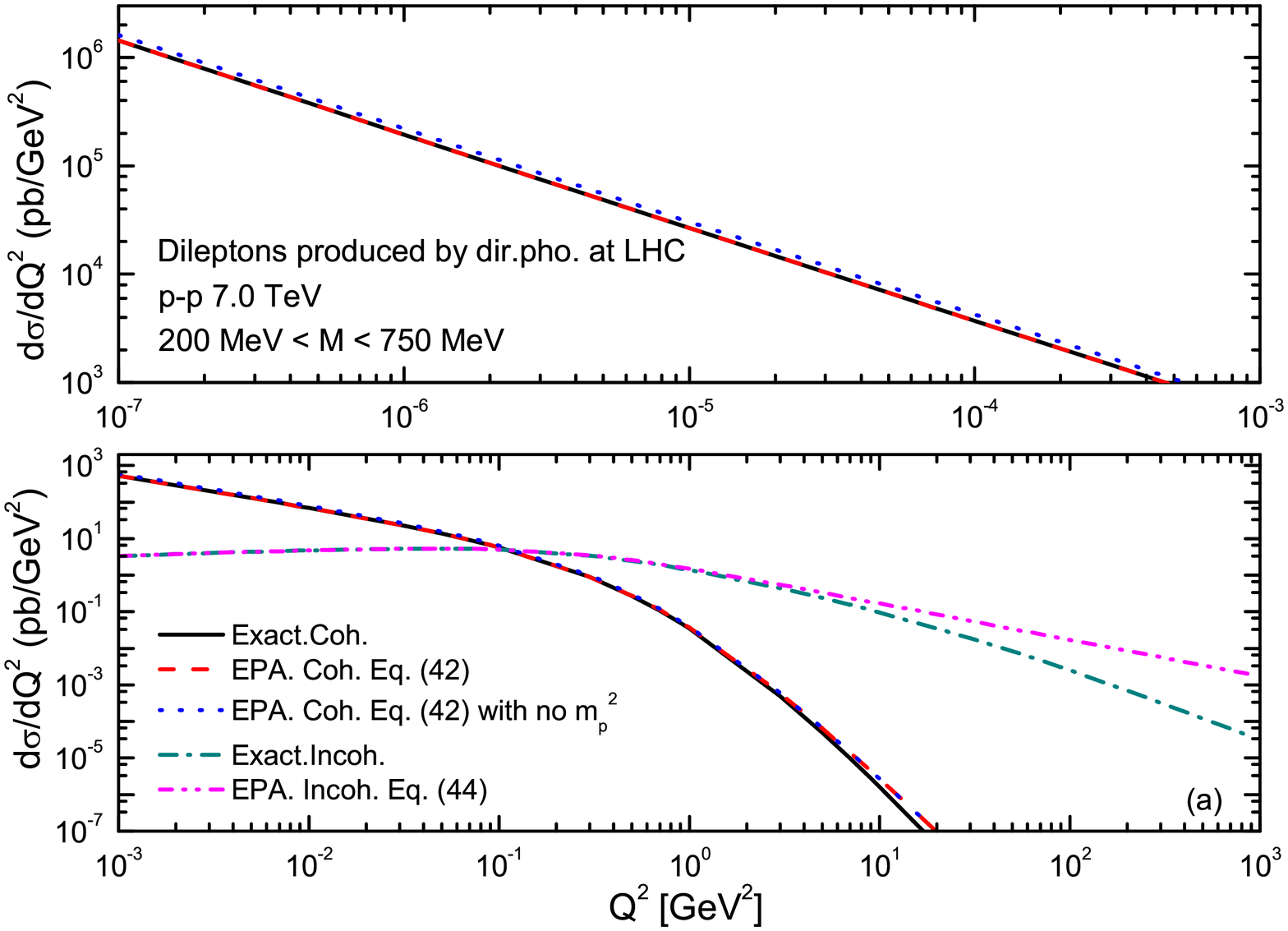}
\end{minipage}
\begin{minipage}[t]{8in}
\includegraphics[height=8.8cm,width=10.3cm]{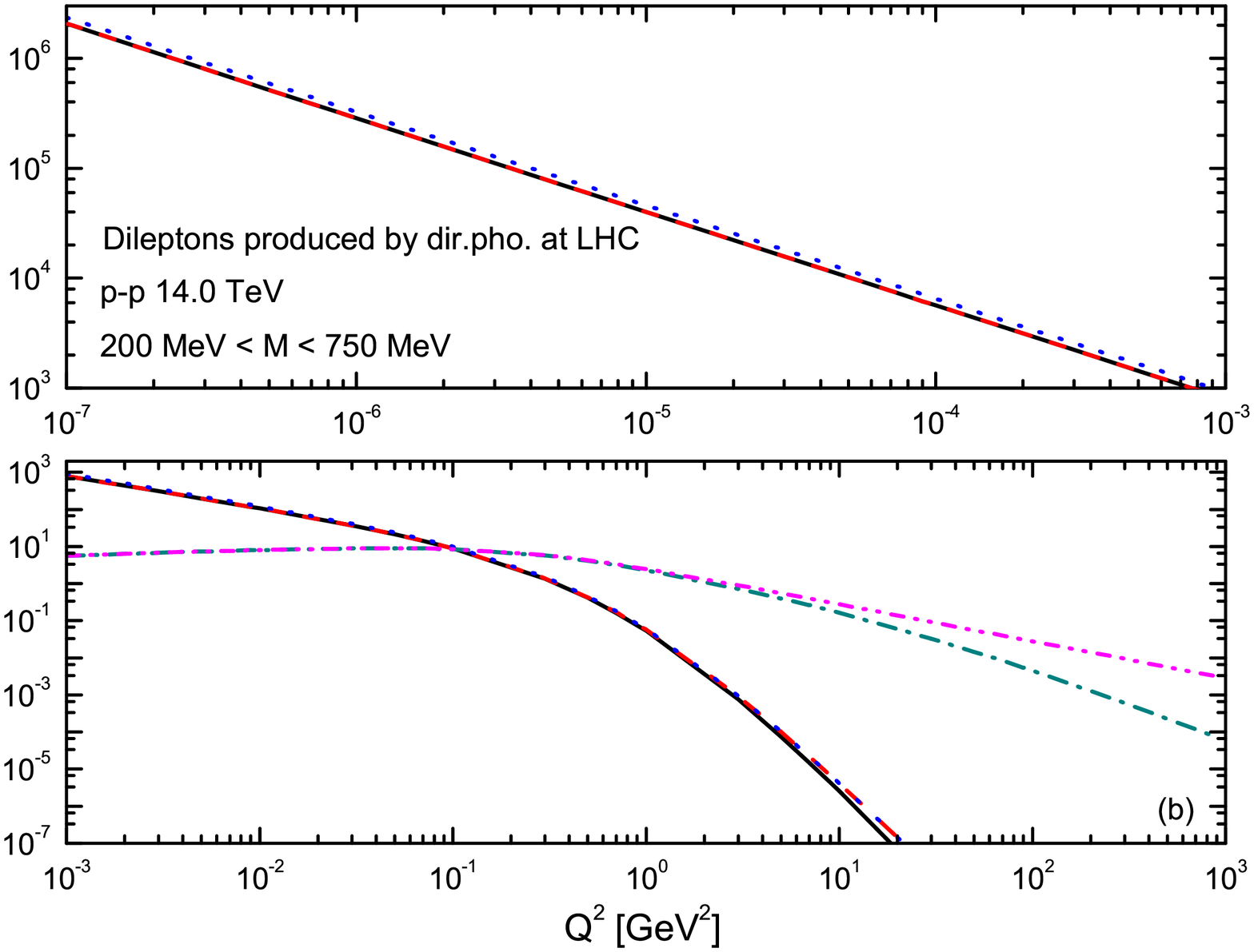}
\end{minipage}
\end{tabular}
\caption{(a) Comparisons of the exact results and EPA ones of dileptons produced by dir.pho for $y_{r}=0$.
 The solid line (black) represents the exact results of coh.pho, the dash line (red) represents the results
of EPA based on the photon flux function Eq. (\ref{fgamma.coh.}), the dot line (blue) represents the results
of EPA based on Eq. (\ref{fgamma.coh.}) but without $m_{p}^{2}$ term, the dash-dot line (dark-cyan) represents
the exact results of incoh.pho, the dash-dot-dot line (magenta) represents the results of EPA based on the
photon flux function of Eq. (\ref{fgamma.incoh.}). (b) The same as figure (a) but for p-p collisions at $\sqrt{s_{NN}}=14.0\ \textrm{TeV}$.
The solid line (black) coincides with the dash line (red) in small $Q^{2}$ domain. Since the contributions of
incoh.pho are so small comparing with coh.pho in the small $Q^{2}$ domain, its results are not plotted in the upper
figures.
}
\label{Q2.dile.dir.}
\end{figure*}
\begin{figure*}
\begin{tabular}{cc}
\begin{minipage}[t]{1.3in}
\includegraphics[height=8.8cm,width=10.3cm]{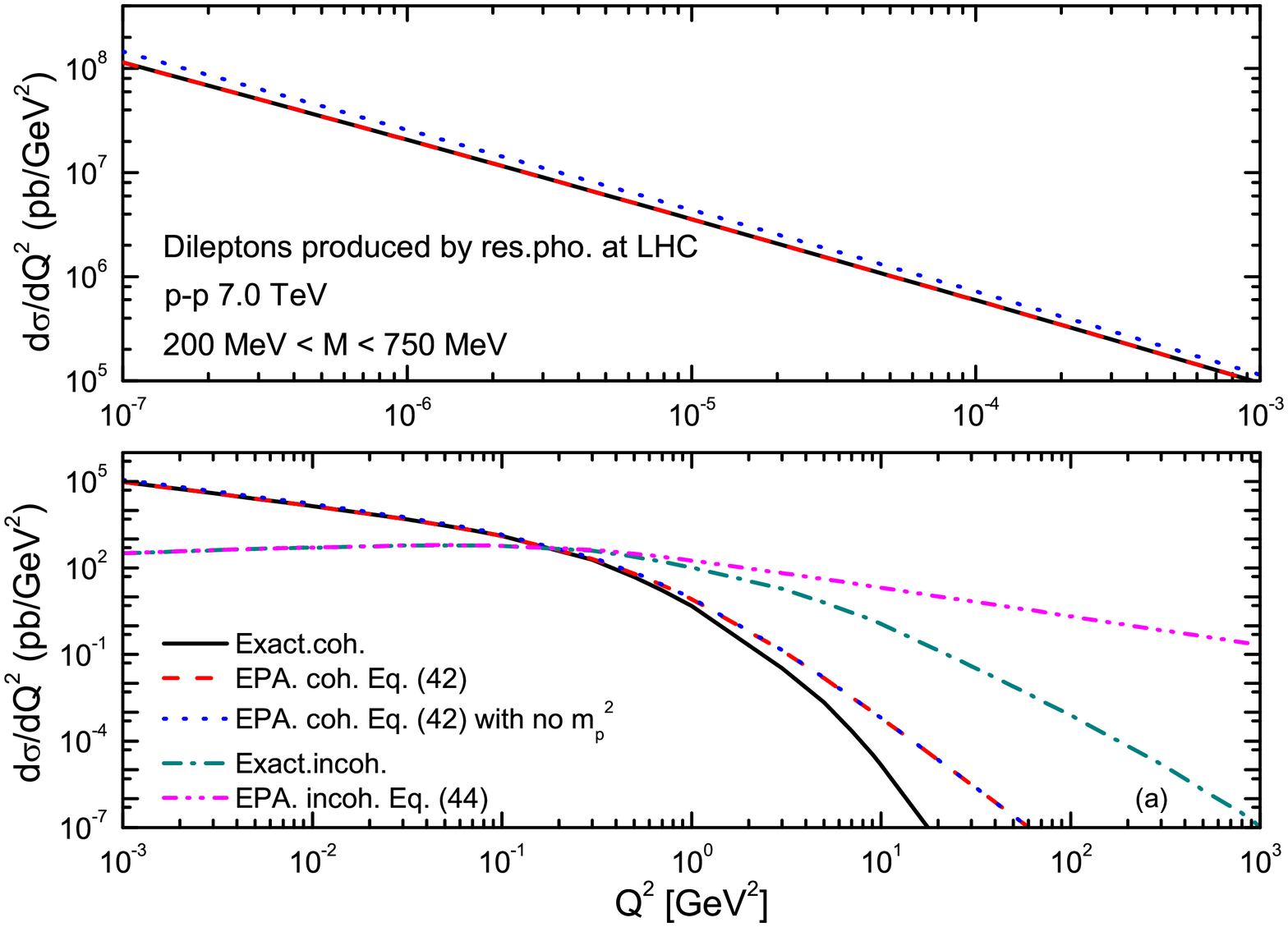}
\end{minipage}
\begin{minipage}[t]{8in}
\includegraphics[height=8.8cm,width=10.3cm]{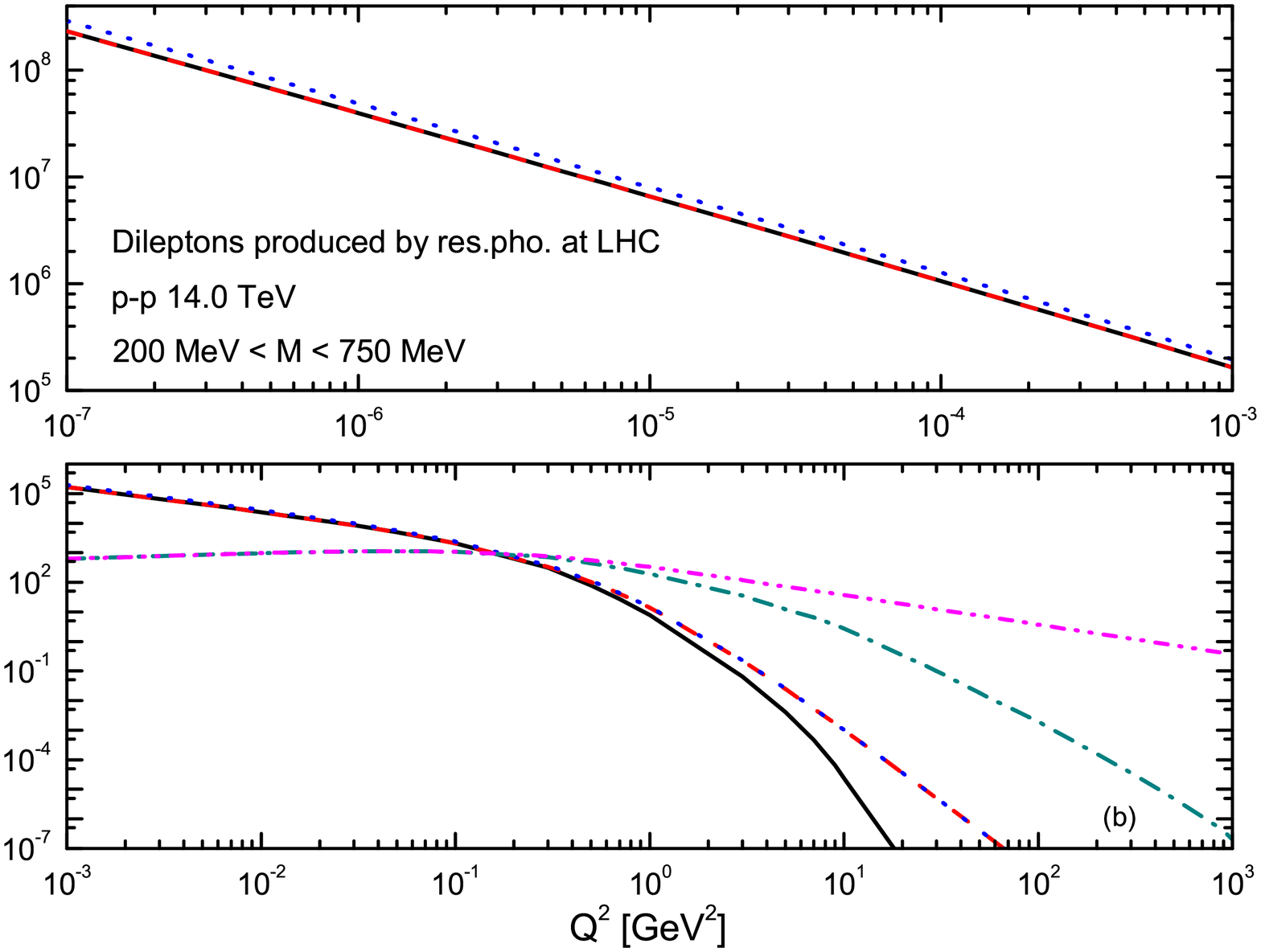}
\end{minipage}
\end{tabular}
\caption{Same as Fig. \ref{Q2.dile.dir.} but for res.pho.
}
\label{Q2.dile.res.}
\end{figure*}
\begin{figure*}
\begin{tabular}{ccc}
\begin{minipage}[t]{5.55cm}
\includegraphics[scale=0.254]{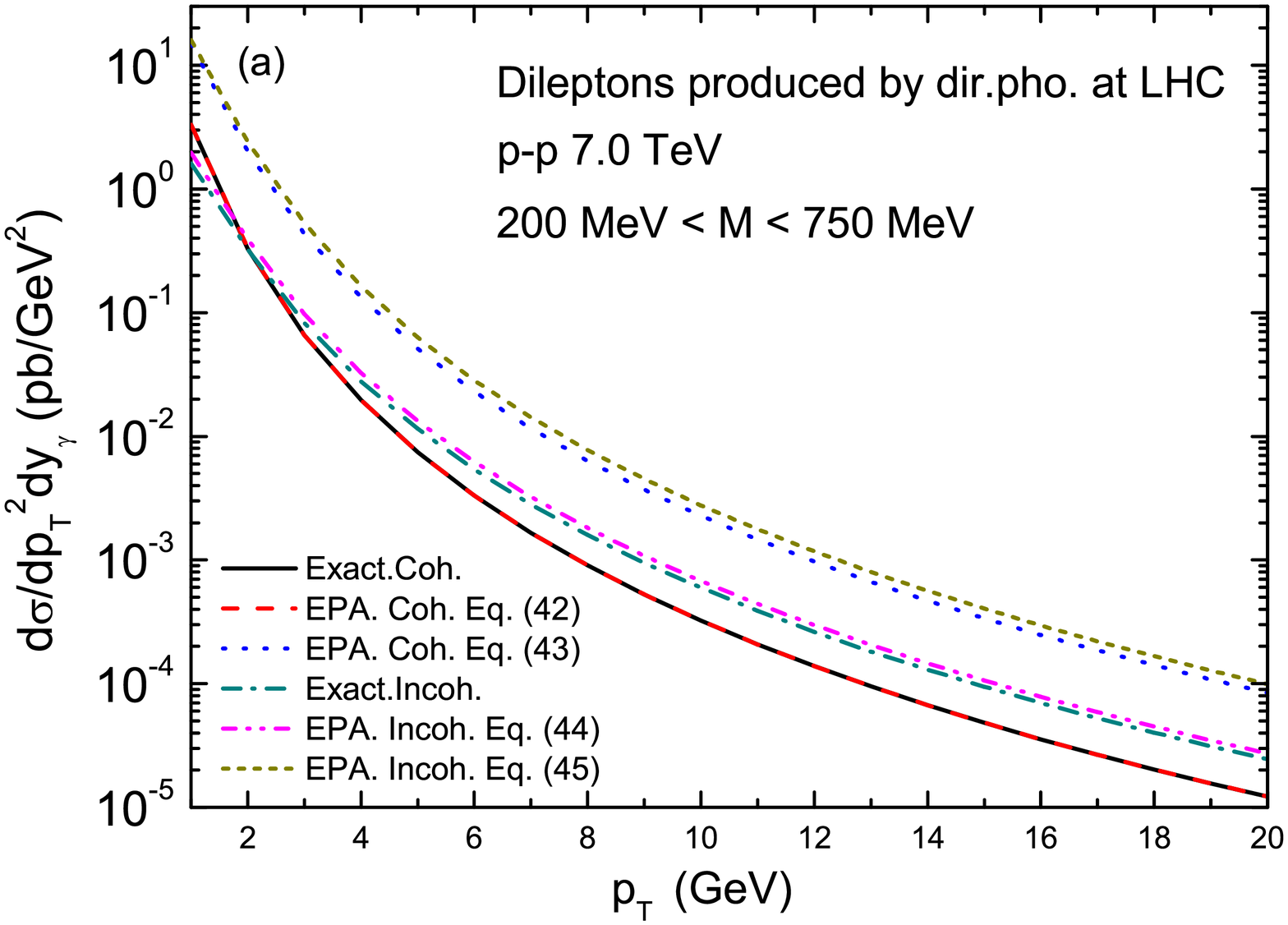}
\end{minipage}
\begin{minipage}[t]{5.3cm}
\includegraphics[scale=0.254]{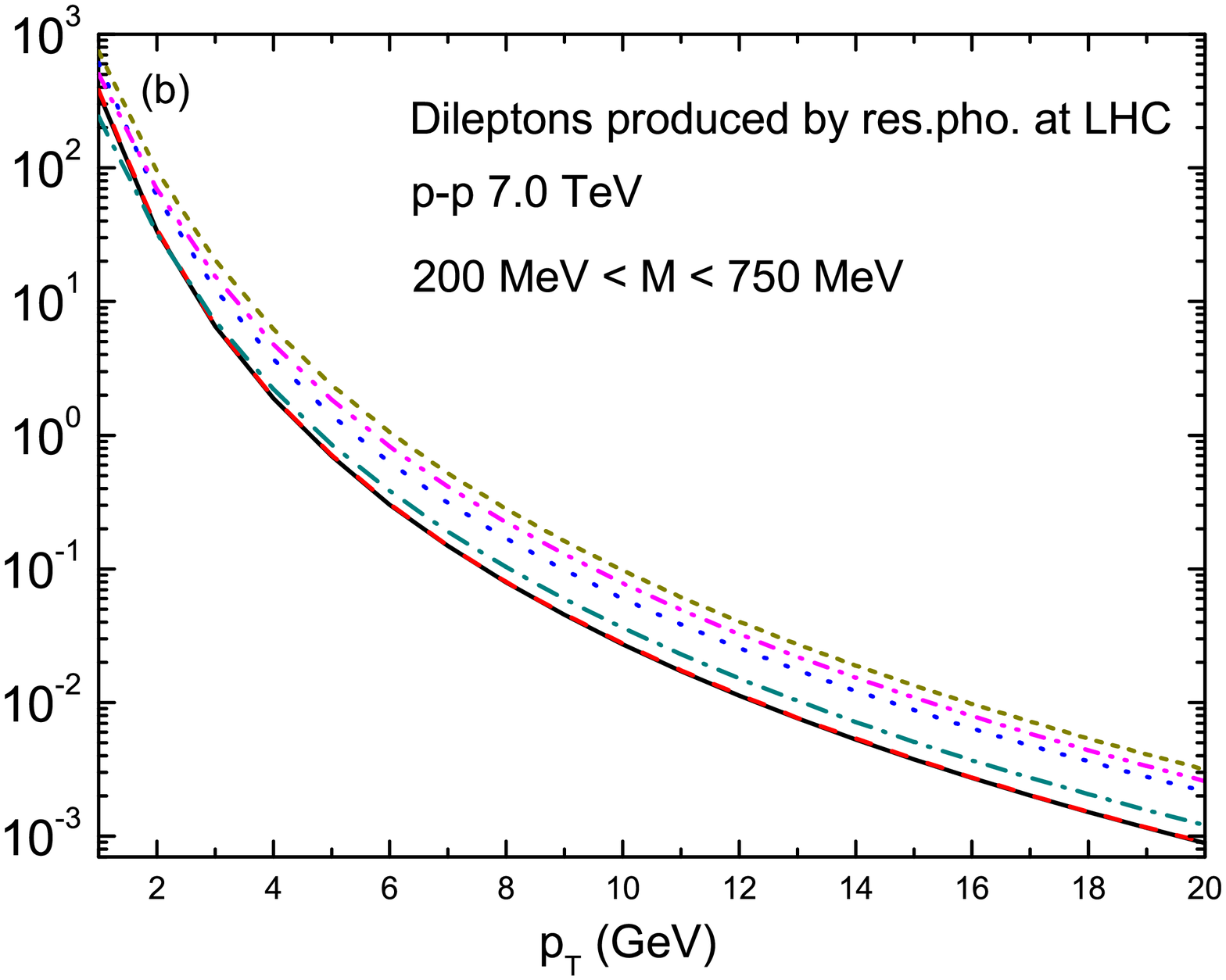}
\end{minipage}
\begin{minipage}[t]{8cm}
\includegraphics[scale=0.254]{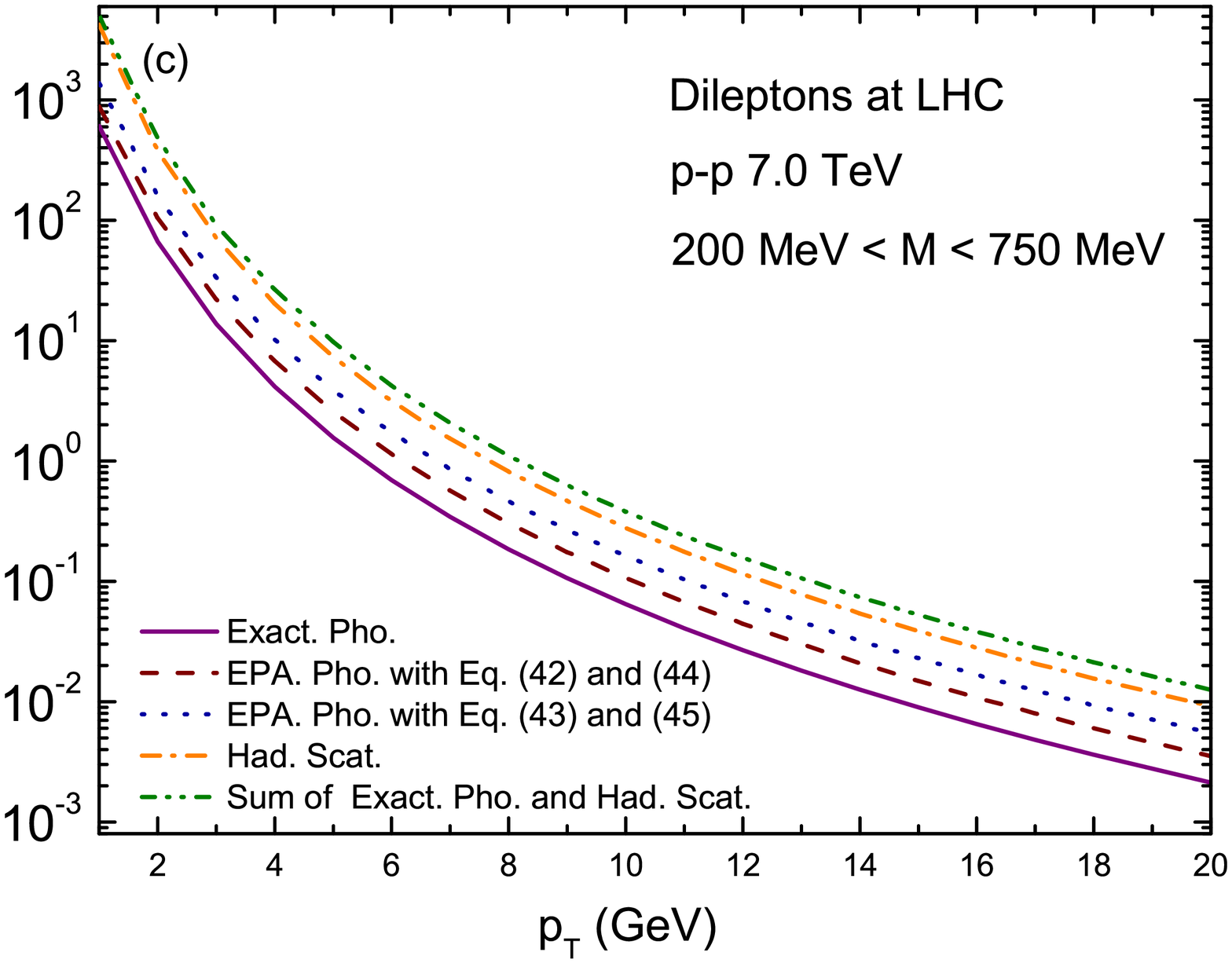}
\end{minipage}
\end{tabular}
\caption{(a) Invariant cross section of dileptons produced by dir.pho. for $y_{r}=0$ in p-p collisions at
$\sqrt{s_{NN}}=7.0\ \textrm{TeV}$. (b) Same as (a) but for res.pho. (c) The comparisons between the photoproduction
processes results with the ones of hadronic processes, the solid line (purple) represents the exact results
of photoproduction processes, the dash line (wine) represents the results of EPA based on Eq. (\ref{fgamma.coh.})
and Eq. (\ref{fgamma.incoh.}), the dot line (royal) represents the results of EPA based on Eq. (\ref{DZ.fgamma.coh.})
and Eq. (\ref{fgamma.incoh.noWF}), the dash-dot line (orange) represents the results of hard scattering of initial
partons (had.scat), the dash-dot-dot line (olive) represents the sum of the exact results of photoproduction
processes and the ones of had.scat. In Fig. 7 (a) and 7 (b), the solid line (black) coincides with the dash
line (red) in the whole $p_{T}$ domain.
}
\label{7PT.dile.}
\end{figure*}
\begin{figure*}
\begin{tabular}{ccc}
\begin{minipage}[t]{5.55cm}
\includegraphics[scale=0.254]{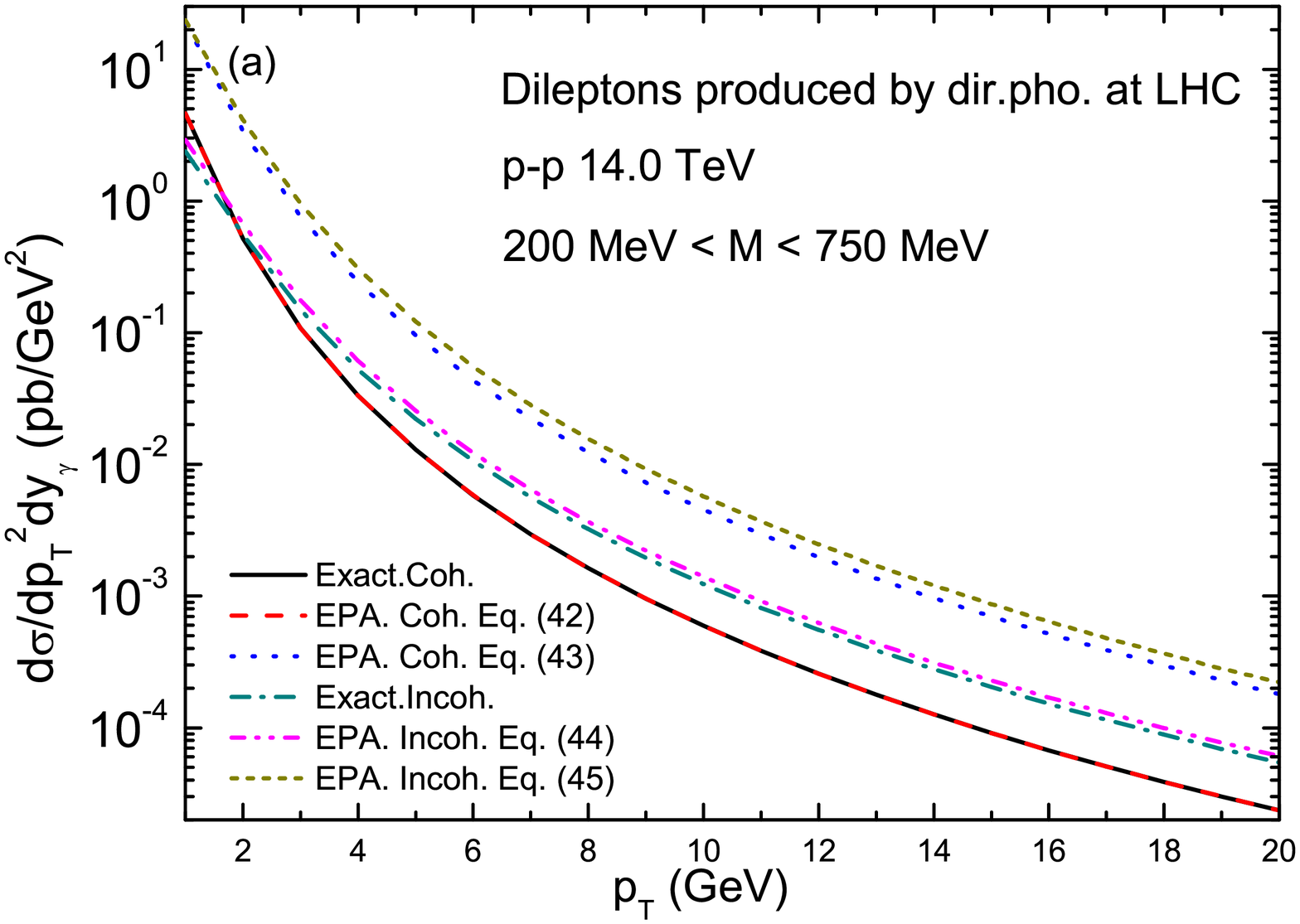}
\end{minipage}
\begin{minipage}[t]{5.3cm}
\includegraphics[scale=0.254]{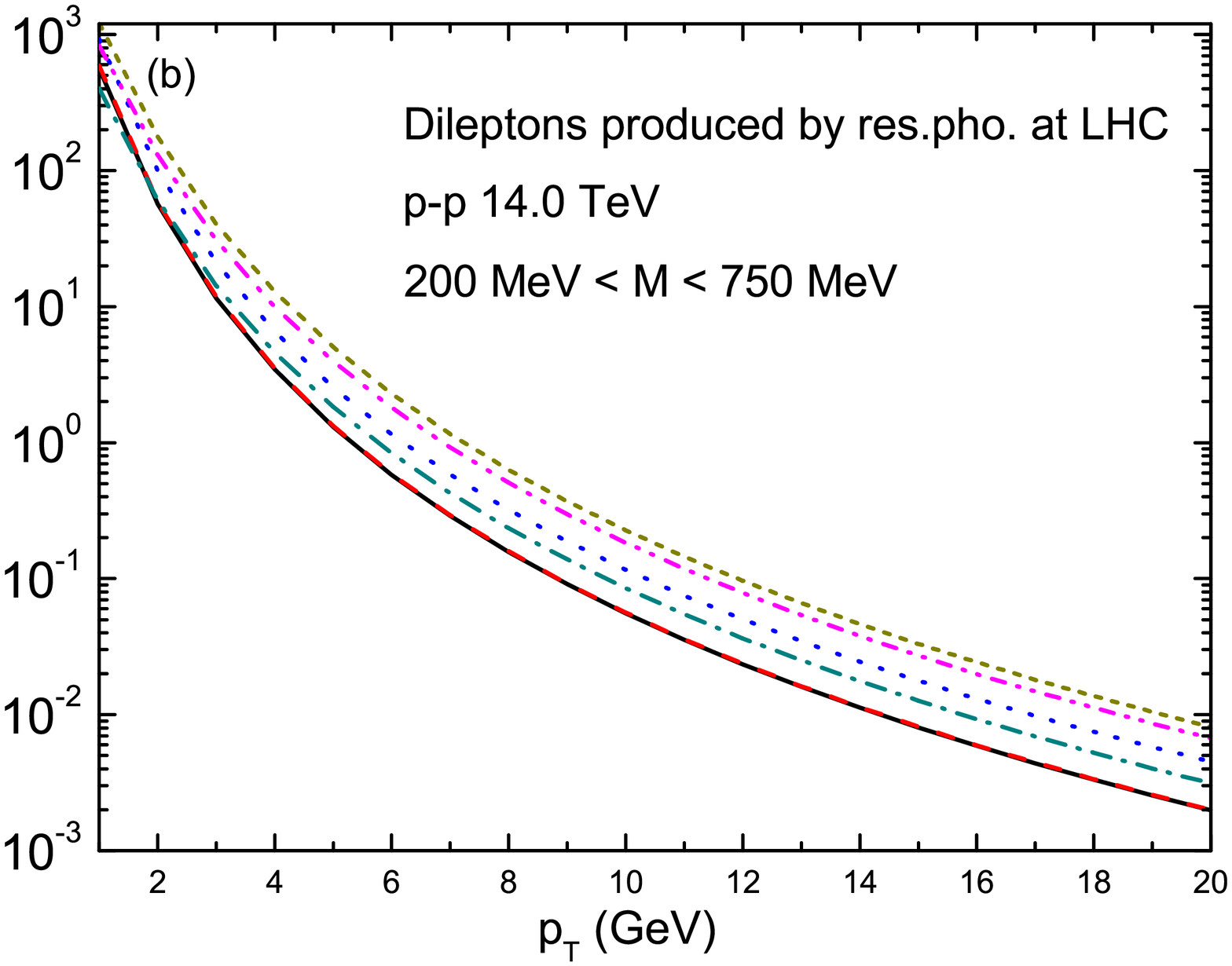}
\end{minipage}
\begin{minipage}[t]{8cm}
\includegraphics[scale=0.254]{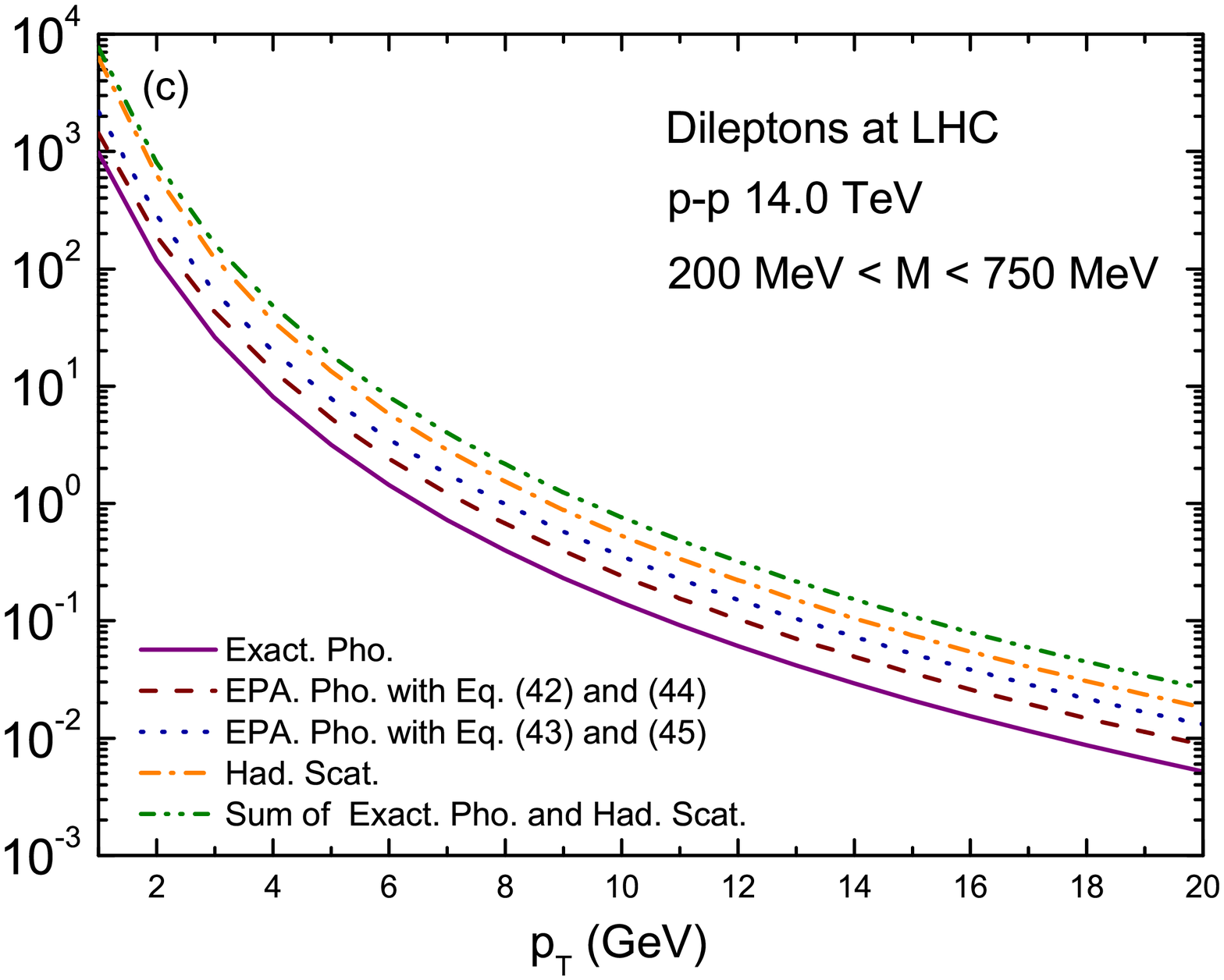}
\end{minipage}
\end{tabular}
\caption{
Same as Fig. \ref{7PT.dile.} but for p-p collisions at $\sqrt{s_{NN}}=14\ \textrm{TeV}$.
}
\label{14PT.dile.}
\end{figure*}
\begin{figure*}
\begin{tabular}{cc}
\begin{minipage}[t]{1.3in}
\includegraphics[height=8.8cm,width=10.3cm]{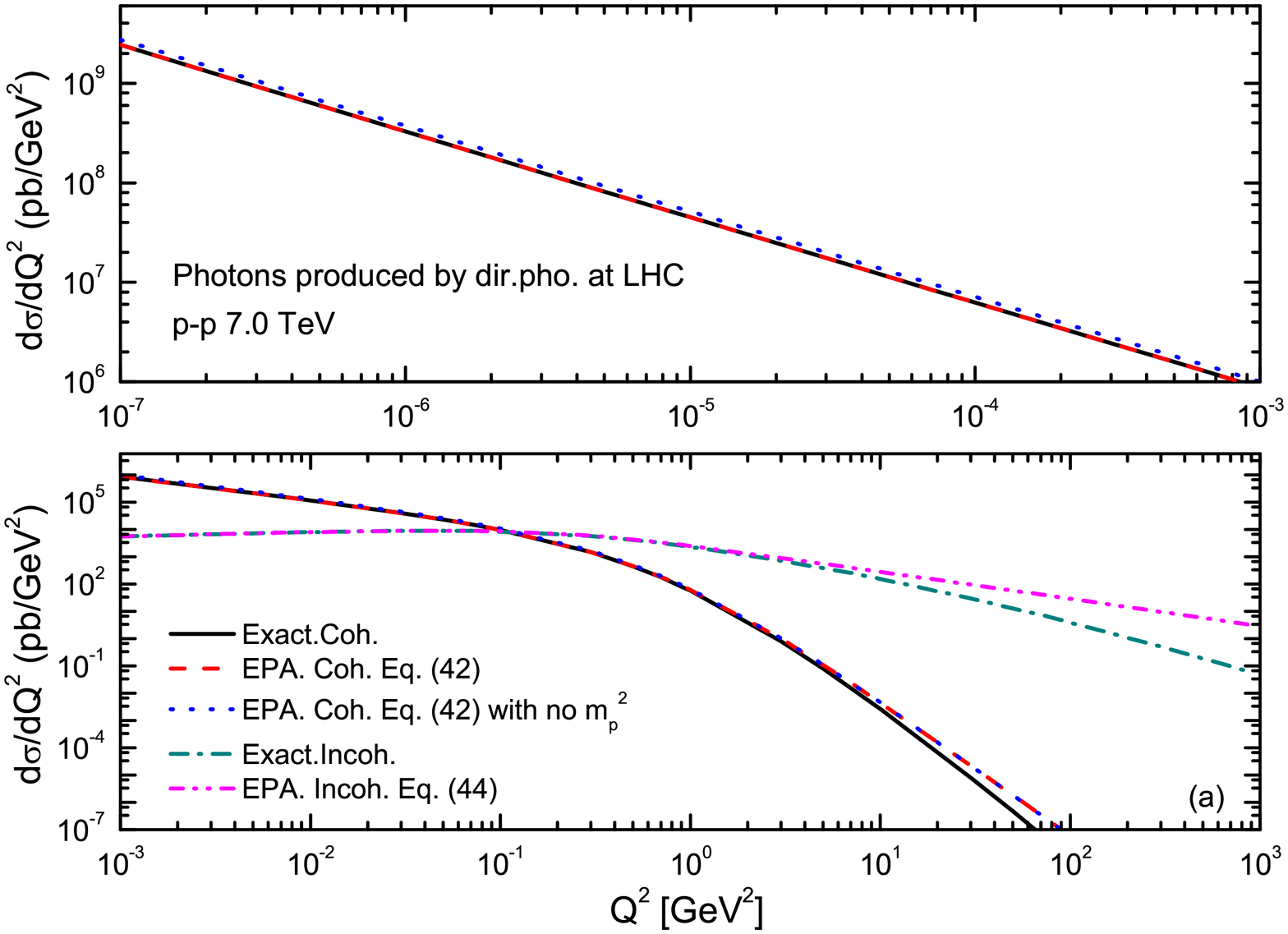}
\end{minipage}
\begin{minipage}[t]{8in}
\includegraphics[height=8.8cm,width=10.3cm]{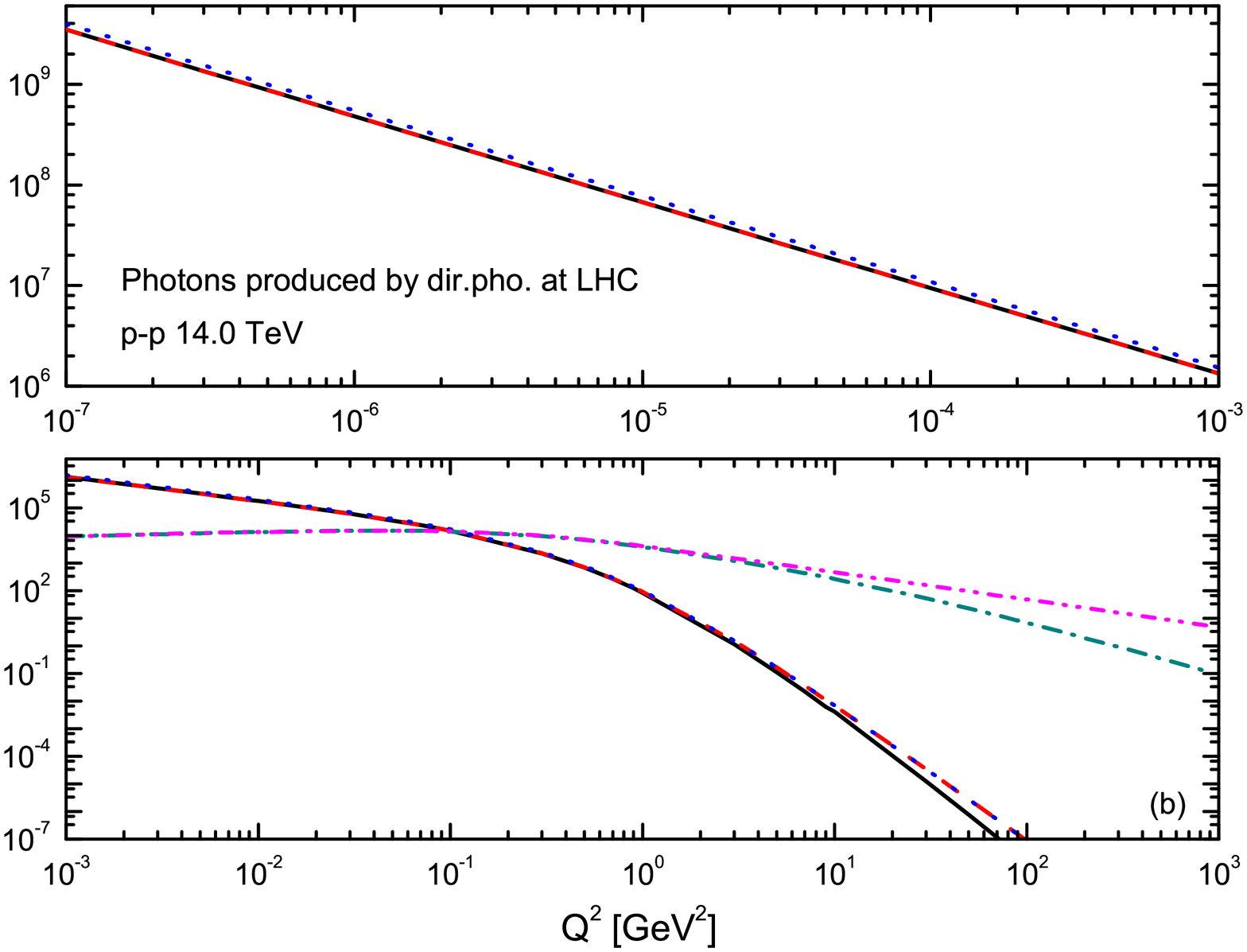}
\end{minipage}
\end{tabular}
\caption{Photons produced by dir.pho. for $y_{r}=0$ in p-p collisions at
$\sqrt{s_{NN}}=7.0\ \textrm{TeV}$ (a) and at\ $\sqrt{s_{NN}}=14.0\
\textrm{TeV}$ (b). The solid line (black) coincides with the dash line (red)
in small $Q^{2}$ domain. Since the contributions of incoh.pho are so small
comparing with coh.pho in the small $Q^{2}$ domain, its results are not
plotted in the upper figures.
}
\label{Q2.pho.dir.}
\end{figure*}
\begin{figure*}
\begin{tabular}{cc}
\begin{minipage}[t]{1.3in}
\includegraphics[height=8.8cm,width=10.3cm]{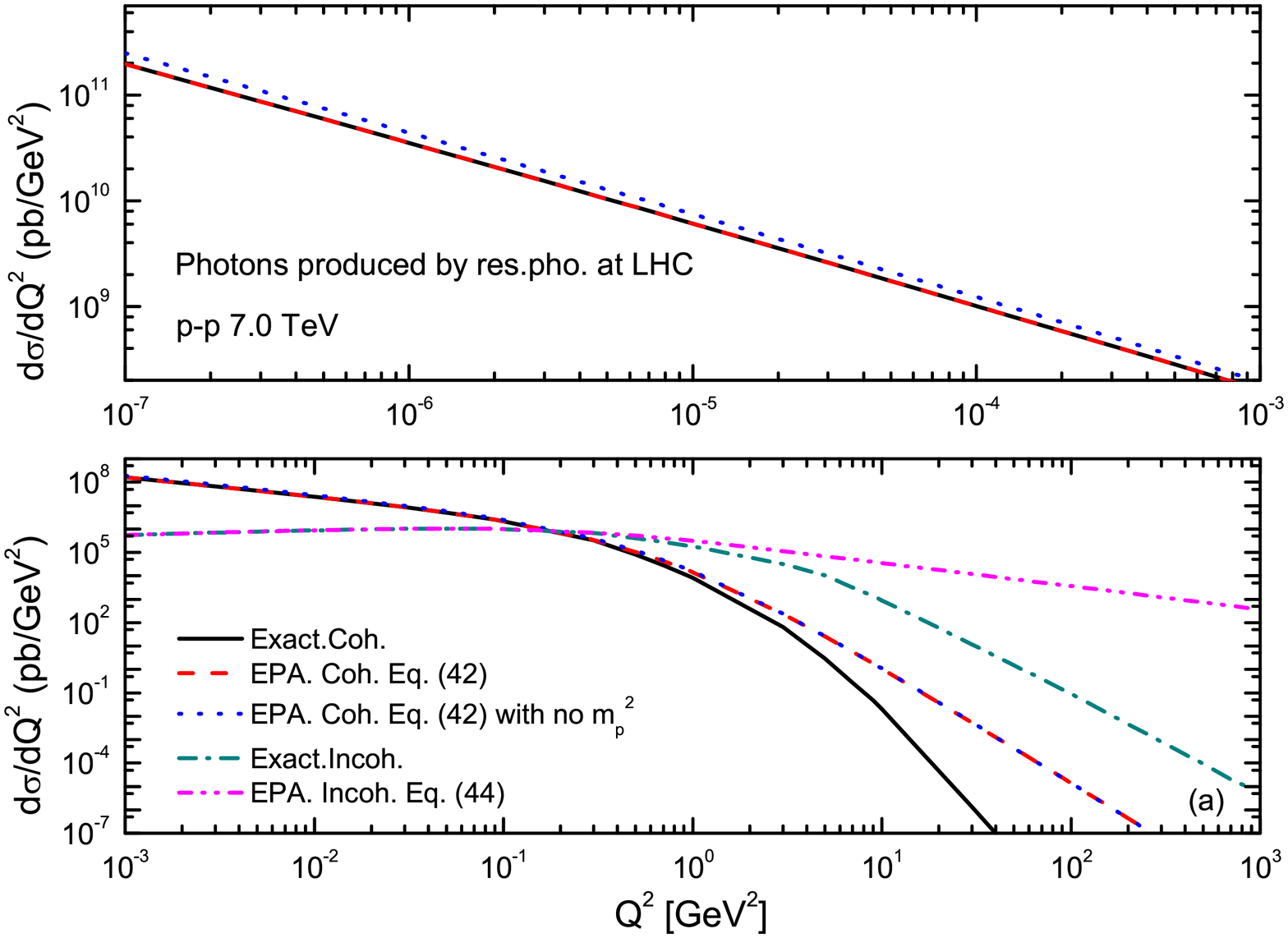}
\end{minipage}
\begin{minipage}[t]{8in}
\includegraphics[height=8.8cm,width=10.3cm]{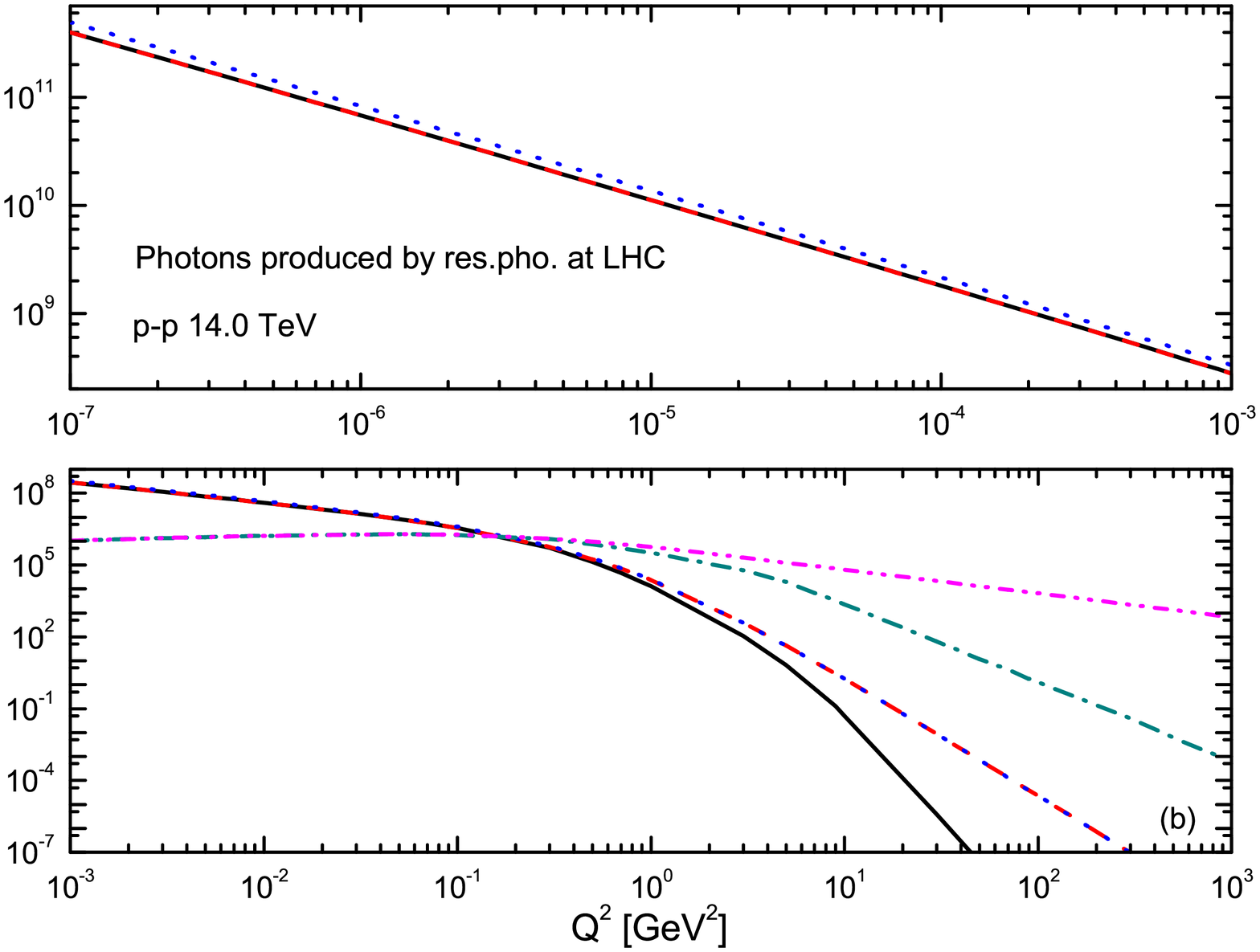}
\end{minipage}
\end{tabular}
\caption{
Same as Fig. \ref{Q2.pho.dir.} but for res.pho.
}
\label{Q2.pho.res.}
\end{figure*}
\begin{figure*}
\begin{tabular}{ccc}
\begin{minipage}[t]{5.55cm}
\includegraphics[scale=0.254]{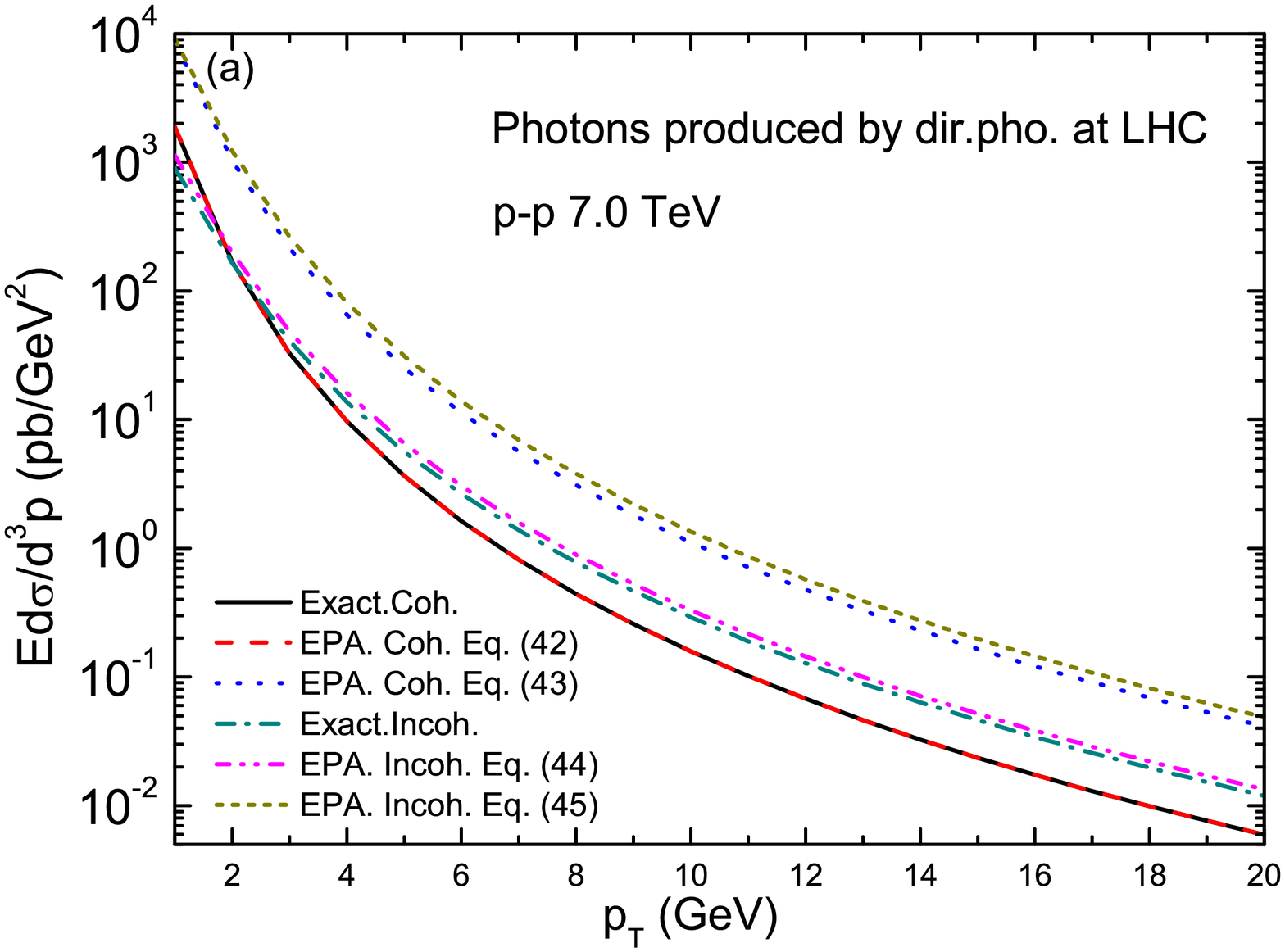}
\end{minipage}
\begin{minipage}[t]{5.3cm}
\includegraphics[scale=0.254]{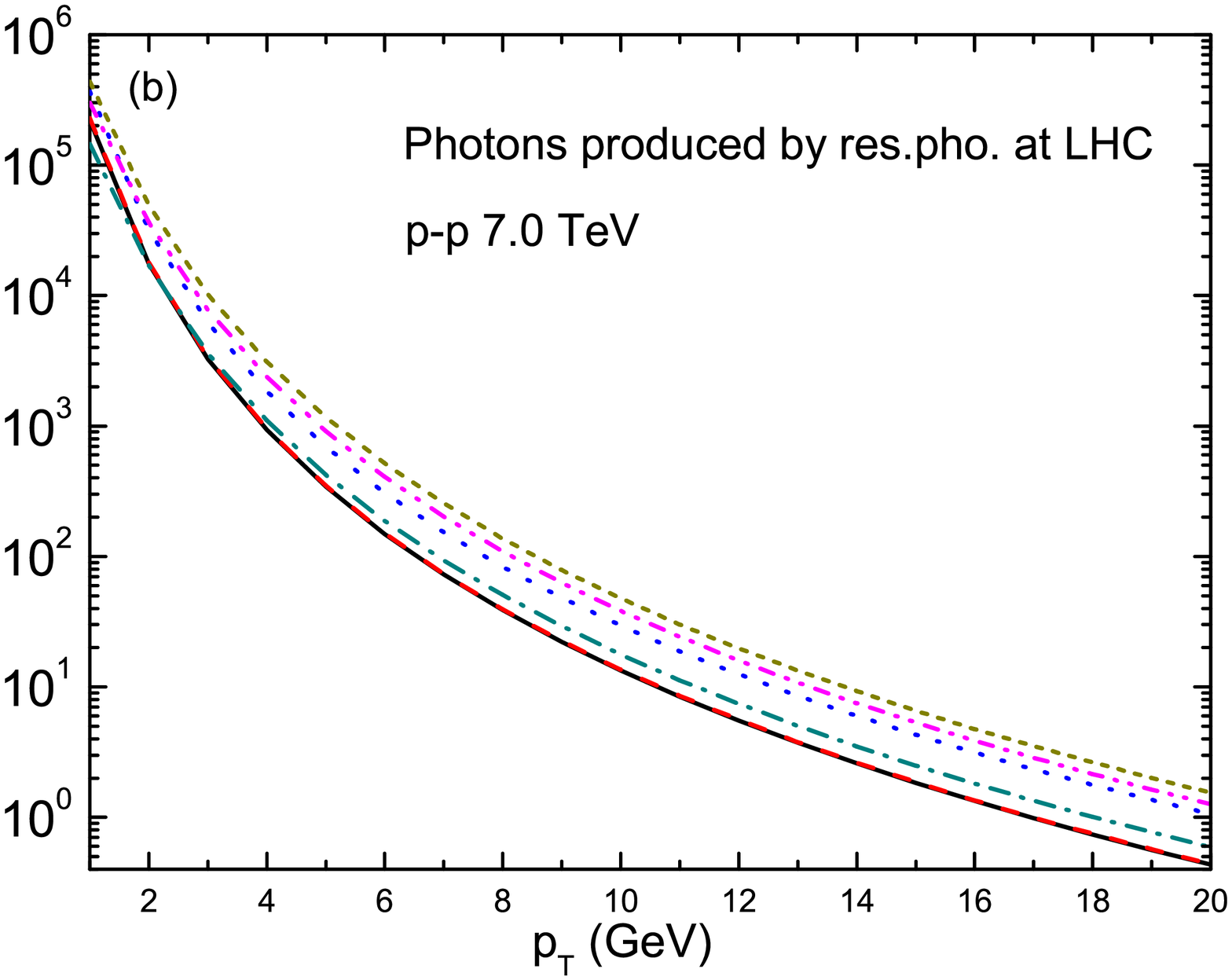}
\end{minipage}
\begin{minipage}[t]{8cm}
\includegraphics[scale=0.254]{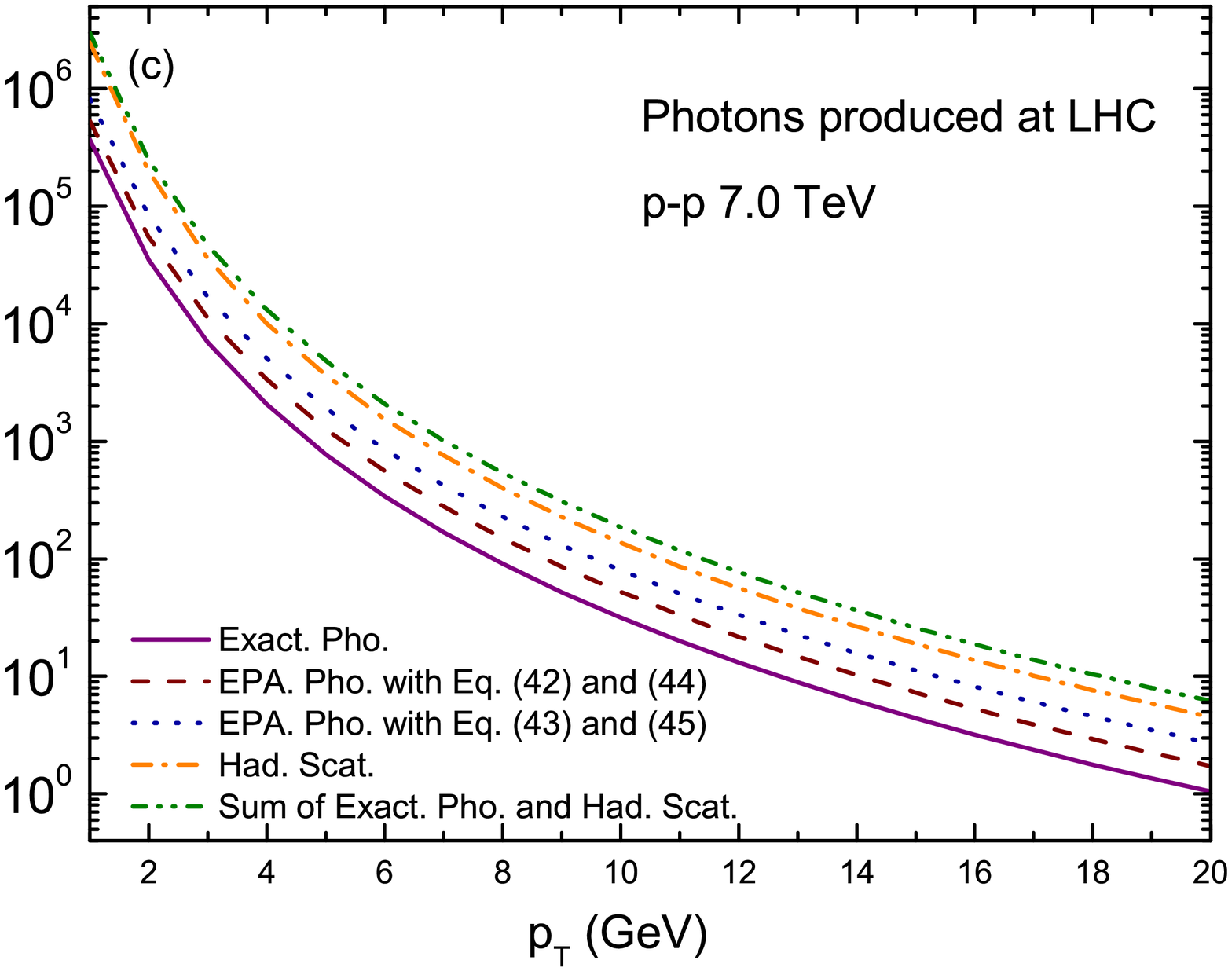}
\end{minipage}
\end{tabular}
\caption{(a) Invariant cross section of photons produced by dir.pho for $y_{r}=0$ in p-p collisions
at $\sqrt{s_{NN}}=7.0\ \textrm{TeV}$. (b) Same as (a) but for res.pho. (c) The comparisons between
the photoproduction processes results with the ones of hadronic processes. In Fig. 11 (a) and 11 (b),
the solid line (black) coincides with the dash line (red) in the whole $p_{T}$ domain.
}
\label{7PT.pho.}
\end{figure*}
\begin{figure*}
\begin{tabular}{ccc}
\begin{minipage}[t]{5.55cm}
\includegraphics[scale=0.254]{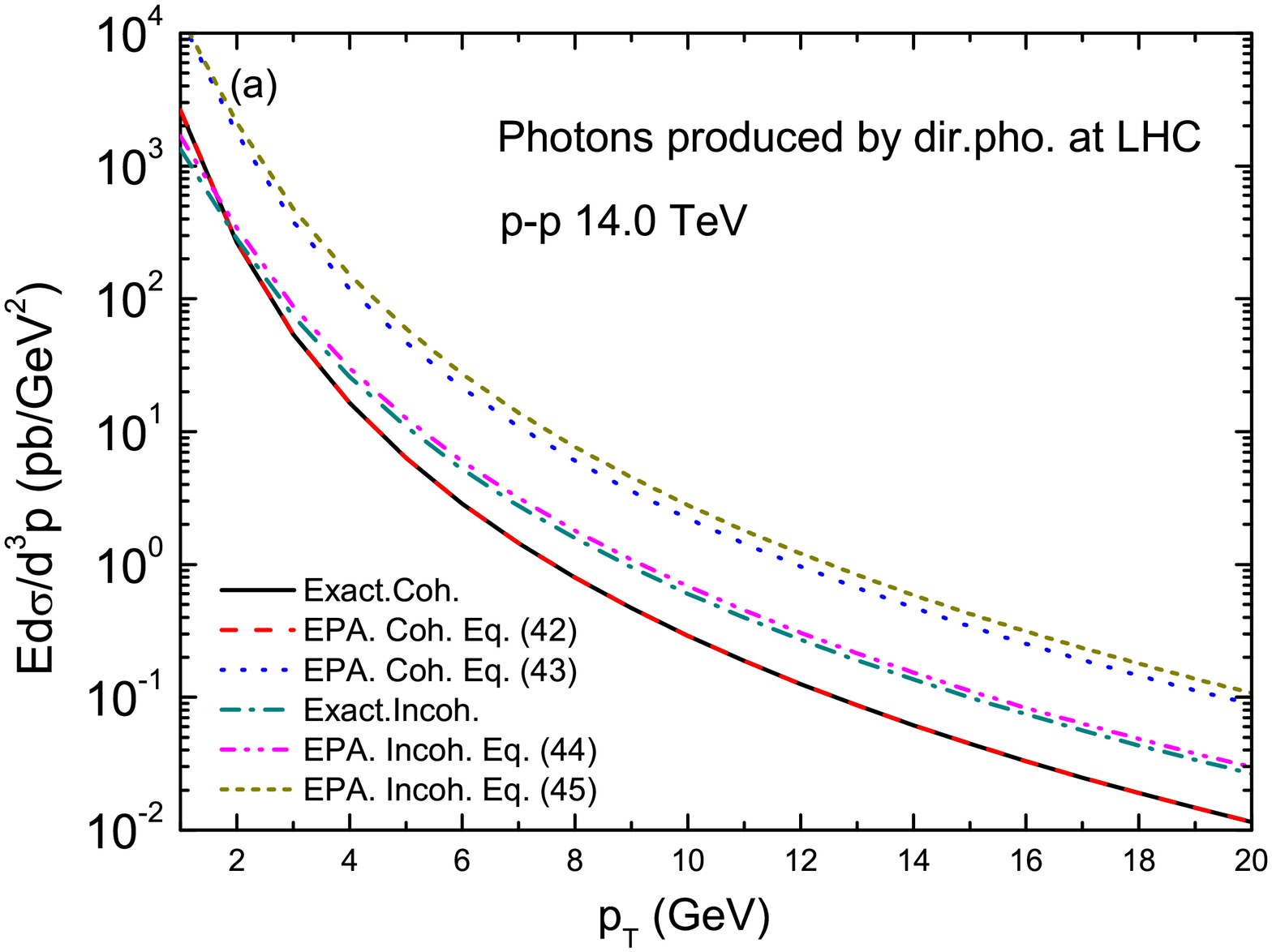}
\end{minipage}
\begin{minipage}[t]{5.3cm}
\includegraphics[scale=0.254]{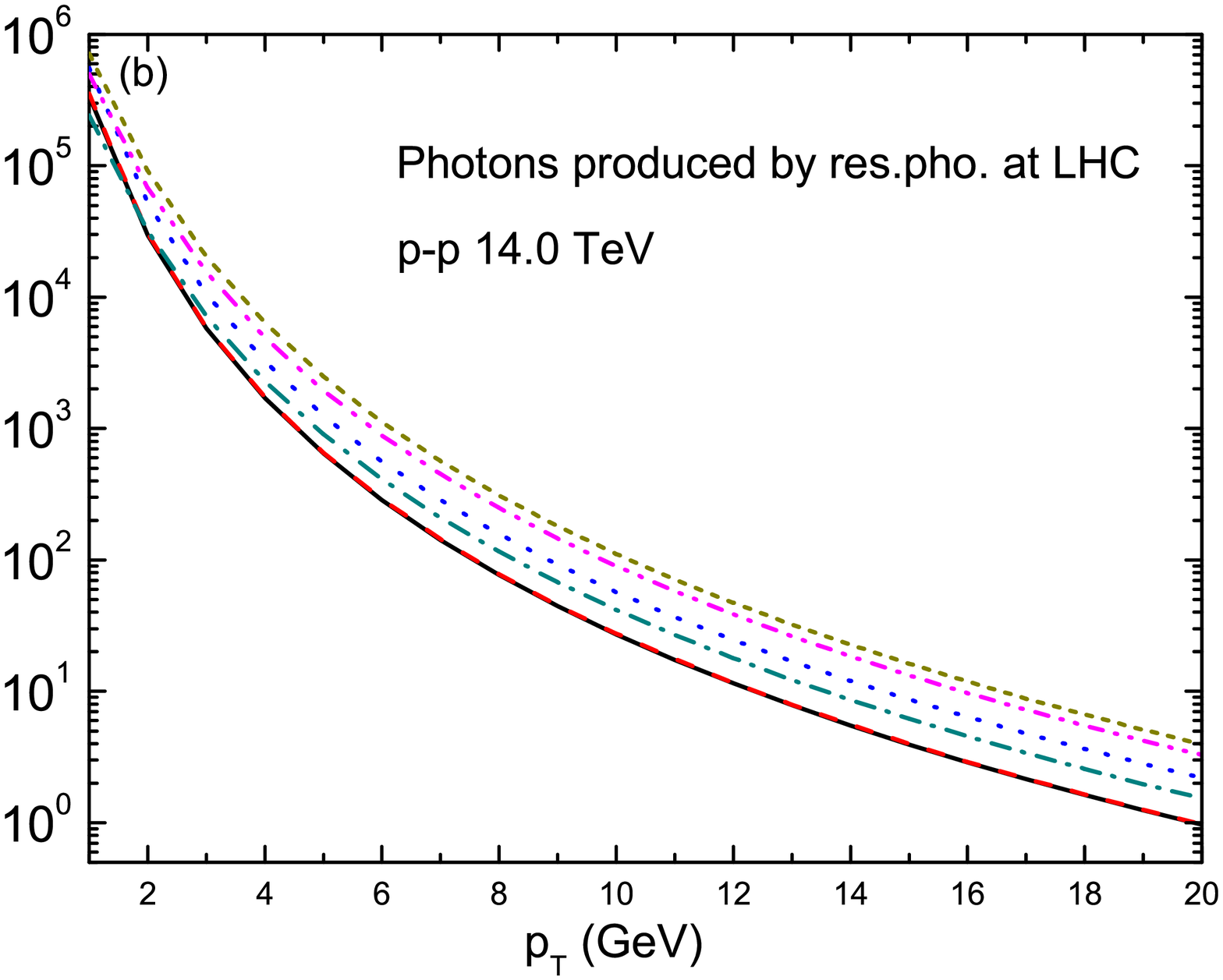}
\end{minipage}
\begin{minipage}[t]{8cm}
\includegraphics[scale=0.254]{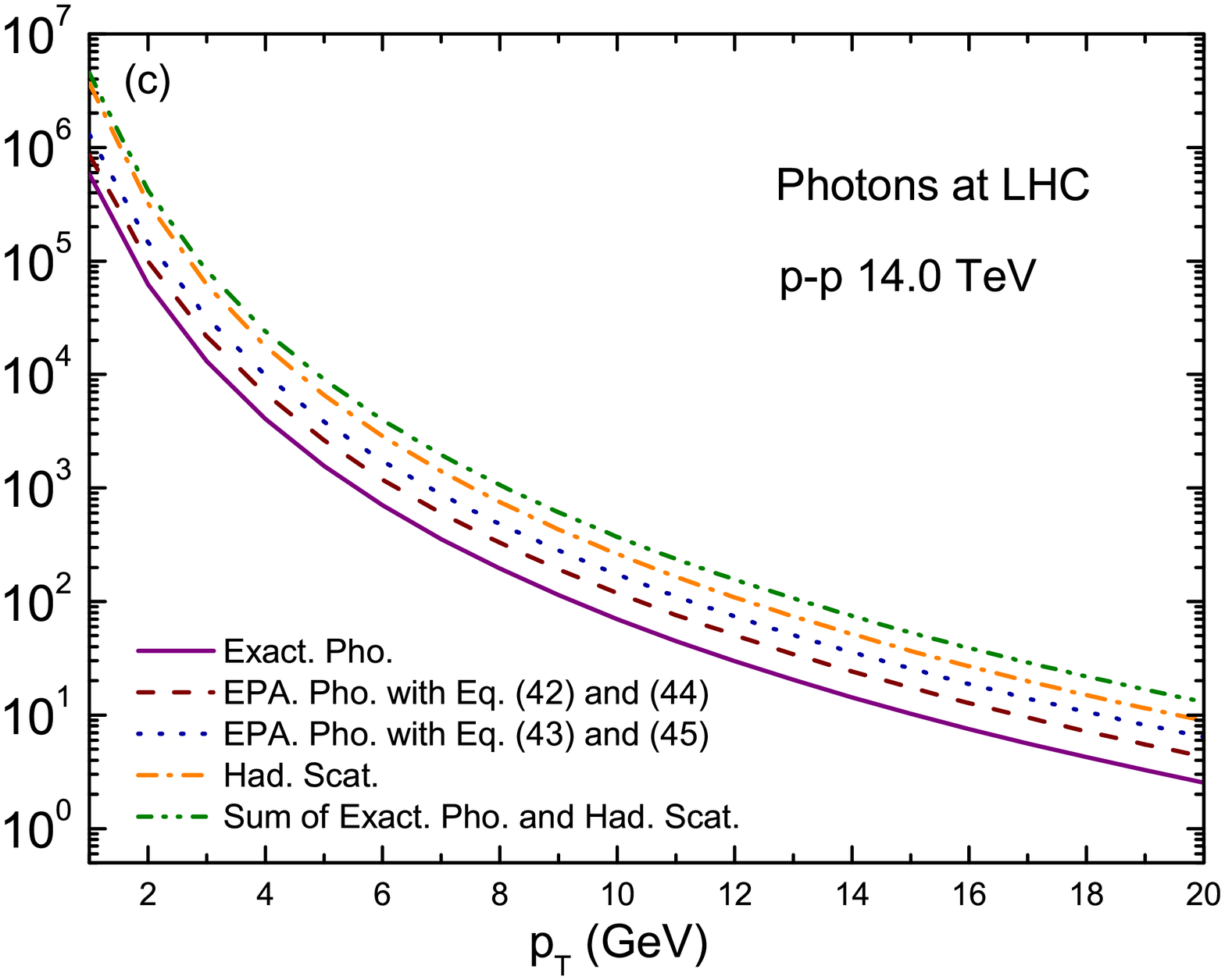}
\end{minipage}
\end{tabular}
\caption{
Same as Fig. \ref{7PT.pho.} but for  p-p collisions at $\sqrt{s_{NN}}=14\ \textrm{TeV}$
}
\label{14PT.pho.}
\end{figure*}
The several theoretical inputs and the bounds of involved variables need to be provided.
The mass range of dileptons is chosen as $200\ \textrm{MeV}<M<750\ \textrm{MeV}$, the mass of
proton is $m_{p}=0.938\ \textrm{GeV}$ \cite{Chin.Phys.C._38_090001}. Since the large $p_{T}$
photoproduction processes are considered (the contribution are mainly from the transverse direction)
, the rapidity is set to be $y_{r}=0$ ($\theta_{c}\sim \pi/2$) according to Ref. \cite{Phys.Rev.C._84_044906}.

For the $Q^{2}$ distribution, the bounds of integration variables for coh.dir are given by
\begin{eqnarray}\label{Inte.Vari.coh.}
&&\hat{t}_{\textrm{min}}=(z_{q\ \textrm{min}}-1)yx_{b}s_{NN},\nonumber\\
&&\hat{t}_{\textrm{max}}=(z_{q\ \textrm{max}}-1)yx_{b}s_{NN},\nonumber\\
&&x_{b\ \textrm{min}}=\frac{\hat{s}_{\textrm{min}}+Q^{2}}{ys_{NN}},\ x_{b\ \textrm{max}}=1,\nonumber\\
&&y_{\textrm{min}}=\frac{\hat{s}_{\textrm{min}}+Q^{2}}{s_{NN}},\nonumber\\
&&y_{\textrm{max}}=\frac{1}{2m_{p}^{2}}\sqrt{4m_{p}^{2}Q^{2}+Q^{4}}+\frac{2m_{p}^{2}-s_{NN}}{2m_{p}^{2}s_{NN}}Q^{2},
\end{eqnarray}
where $\hat{s}_{\textrm{min}}=(M_{T\ \textrm{min}}+p_{T\ \textrm{min}})^{2}$, $p_{T}^{2}=\hat{t}(\hat{s}\hat{u}+Q^{2}M^{2})/(\hat{s}+Q^{2})^{2}$
is the square of the transverse momentum for dileptons and
\begin{eqnarray}\label{z.coh.dir}
&&z_{q\ \textrm{min}}=\frac{M^{2}+\hat{s}}{2\hat{s}}-\frac{\sqrt{(\hat{s}-M^{2})^{2}-4p_{T\ \textrm{min}}^{2}\hat{s}}}{2\hat{s}}\nonumber\\
&&z_{q\ \textrm{max}}=\frac{M^{2}+\hat{s}}{2\hat{s}}+\frac{\sqrt{(\hat{s}-M^{2})^{2}-4p_{T\ \textrm{min}}^{2}\hat{s}}}{2\hat{s}}.
\end{eqnarray}

The bounds of variables for the incoh.dir are same as Eq. (\ref{Inte.Vari.coh.}), but for $\hat{t}_{\textrm{min}}=
(z_{q\ \textrm{min}}-1)yx_{a}x_{b}s_{NN}$, $\hat{t}_{\textrm{max}}=(z_{q\ \textrm{max}}-1)yx_{a}x_{b}s_{NN}$,
$x_{b\ \textrm{min}}=(\hat{s}_{\textrm{min}}+Q^{2})/yx_{a}s_{NN}$,
$x_{a\ \textrm{min}}=(\hat{s}_{\textrm{min}}+Q^{2})/ys_{NN}$ and $x_{a\ \textrm{max}}=1$.

In the coh.res, the bounds of variables are
\begin{eqnarray}\label{Vari.cohres}
&&\hat{t}_{\gamma\ \textrm{min}}=(z_{q\ \textrm{min}}'-1)z_{a}yx_{b}s_{NN},\nonumber\\
&&\hat{t}_{\gamma\ \textrm{max}}=(z_{q\ \textrm{max}}'-1)z_{a}yx_{b}s_{NN},\nonumber\\
&&z_{a\ \textrm{min}}=\frac{\hat{s}_{\gamma\ \textrm{min}}}{yx_{b}s_{NN}},\ z_{a\ \textrm{max}}=1,\nonumber\\
&&x_{b\ \textrm{min}}=\frac{\hat{s}_{\gamma\ \textrm{min}}}{z_{a\ \textrm{max}}ys_{NN}},\ x_{b\ \textrm{max}}=1,\nonumber\\
&&y_{\textrm{min}}=\frac{\hat{s}_{\gamma\ \textrm{min}}}{z_{a\ \textrm{max}}s_{NN}},
\end{eqnarray}
$y_{\textrm{max}}$ is same as Eq. (\ref{Inte.Vari.coh.}), where $\hat{s}_{\gamma\ \textrm{min}}=(M_{T\ \textrm{min}}+p_{T\ \textrm{min}})^{2}$, $p_{T}^{2}=\hat{t}\hat{u}/\hat{s}$ and
\begin{eqnarray}\label{z.coh.dir}
&&z_{q\ \textrm{min}}'=\frac{M^{2}+\hat{s}_{\gamma}}{2\hat{s}_{\gamma}}-\frac{\sqrt{(\hat{s}_{\gamma}-M^{2})^{2}-4p_{T\ \textrm{min}}^{2}\hat{s}_{\gamma}}}{2\hat{s}_{\gamma}}\nonumber\\
&&z_{q\ \textrm{max}}'=\frac{M^{2}+\hat{s}_{\gamma}}{2\hat{s}_{\gamma}}+\frac{\sqrt{(\hat{s}_{\gamma}-M^{2})^{2}-4p_{T\ \textrm{min}}^{2}\hat{s}_{\gamma}}}{2\hat{s}_{\gamma}}.\nonumber\\
\end{eqnarray}

The bounds of variables for the incoh.res are same as Eq. (\ref{Vari.cohres})
but for $\hat{t}_{\gamma\ \textrm{min}}=(z_{q\ \textrm{min}}'-1)z_{a}yx_{a}x_{b}s_{NN}$,
$\hat{t}_{\gamma\ \textrm{max}}=(z_{q\ \textrm{max}}'-1)z_{a}yx_{a}x_{b}s_{NN}$, $z_{a\
\textrm{min}}=\hat{s}_{\gamma\ \textrm{min}}/(yx_{a}x_{b}s_{NN})$, $x_{b\ \textrm{min}}=
\hat{s}_{\gamma\ \textrm{min}}/(z_{a\ \textrm{max}}yx_{a}s_{NN})$, $x_{a\ \textrm{min}}=
\hat{s}_{\gamma\ \textrm{min}}/(z_{a\ \textrm{max}}ys_{NN})$ and $x_{a\ \textrm{max}}=1$.

For the $p_{T}$ distribution, the bounds of $x_{a}$, $x_{b}$ and $y$ are same as
$Q^{2}$ distribution, but for $z_{a\ \textrm{max}}=1/(1+Q^{2}/(4p_{T}^{2}))$ \cite{Phys.Rev.D_29_852, Phys.Rev.D._51_3320},
the bounds of $Q^{2}$ are $Q^{2}_{\textrm{min}|\textrm{coh.dir}}=x_{1}^{2}m_{p}^{2}/(1-x_{1})$, $x_{1}=\hat{s}/s_{NN}$, $Q^{2}_{\textrm{min}|\textrm{incoh.dir}}=0$, $Q^{2}_{\textrm{min}|\textrm{coh.res}}=Q^{2}_{\textrm{min}|\textrm{incoh.res}}=0.01\ \textrm{GeV}^{2}$ \cite{EPJC_10_313}, and $Q^{2}_{\textrm{max}}=4p_{T}^{2}$ is used for the exact calculations
and the EPA ones Eq. (\ref{fgamma.coh.}) and Eq. (\ref{fgamma.incoh.}).

In Fig. 5 and 6, the $Q^{2}$ distribution of dileptons produced by photoproduction processes
in p-p collisions at LHC energies are plotted. The contribution of exact treatment are compared
with the EPA ones. For the case of coh.dir, the results of EPA share the same trend with the
exact one in the small $Q^{2}$ region, since EPA is obtained by setting the photon virtuality
$Q^{2}\rightarrow 0$ and neglecting the longitudinal photon contributions. Considering that
the coherent photon flux function with the DZ form Eq. (\ref{DZ.fgamma.coh.}) is obtained by
neglecting the $m_{p}^{2}$ term, the EPA result Eq. (\ref{fgamma.coh.}) with no $m_{p}^{2}$ term
is also presented for researching the $Q^{2}$ dependence behaviour and the validity of Eq. (\ref{DZ.fgamma.coh.}).
It can be seen that, the EPA result with no $m_{p}^{2}$ term is greater than the result
of Eq. (\ref{fgamma.coh.}) at small $Q^{2}$ domain, but they become consistent with increasing
$Q^{2}$, since the $m_{p}^{2}$ term is inversely proportional to $Q^{2}$. The exact result
is in agreement with the EPA results Eq. (\ref{fgamma.coh.}) in small $Q^{2}$ region, and is less than the EPA ones
when $Q^{2}>10\ \textrm{GeV}^{2}$. The case of coh.res is similar to coh.dir, but the
differences between the exact results and the EPA ones are much more evident in large $Q^{2}$
domain. Therefore, the EPA approach is only suitable in the small $Q^{2}$ domain, and can be
used as a good approximation for coh.pho, since the small $Q^{2}$ domain give the main
contribution which agree with the statements of Martin and Ryskin in Ref. \cite{EPJC_74_3040},
and of Budnev and Ginzburg in Ref. \cite{Phys.Rep._15_181}. And the errors from the omission
of $m_{p}^{2}$ term in Eq. (\ref{fgamma.coh.}) and the option of $Q^{2}_{\textrm{max}}\sim\hat{s}\ or\ \infty$
in Eq. (\ref{fgamma.coh.}) and Eq. (\ref{DZ.fgamma.coh.}) can not be neglected.

For the case of incoh.dir, the exact result and the EPA ones are almost same
and can be neglected comparing with coh.dir when $Q^{2}<0.01\ \textrm{GeV}^{2}$,
but the differences among them are evident when $Q^{2}>0.1\ \textrm{GeV}^{2}$.
Besides, the incoh.dir contribution is comparable with the coh.dir contribution
when $Q^{2}>0.01\ \textrm{GeV}^{2}$, and becomes much larger when $Q^{2}>0.1\ \textrm{GeV}^{2}$.
The case of incoh.res is similar to incoh.dir, but the differences between the
exact results and the EPA ones are more prominent in large $Q^{2}$ domain.
It should be emphasized that, if the Martin-Ryskin method is not considered, the
incoh.pho contribution will always much larger than the coh.pho one in the whole
$Q^{2}$ region and is divergent at very small $Q^{2}$ domain ($Q^{2}\rightarrow 0$).
This is a unphysical result. Comparing with the Martin-Ryskin method which avoid
this unphysical large value of incoh.pho naturally, the physical interpretation
is not clear in literatures \cite{Nucl.Phys.B._900_431, Phys.Rev.C._92_054907, Phys.Rev.C._84_044906,
Phys.Rev.C._91_044908, Chin.Phys.Lett._29_081301} which calculated the incoh.pho contribution
by using the artificial cutoff $Q^{2}>1\ \textrm{GeV}^{2}$. Therefore, the EPA approach
is not a effective approximation for incoh.pho, since the incoh.pho contribution are mainly
from the large $Q^{2}$ domain where the errors are obvious, and the results in these
works are not accurate enough.

The $p_{T}$ distribution of dileptons produced by photoproduction processes in p-p
collisions at LHC energies are illustrated in Fig. \ref{7PT.dile.} and \ref{14PT.dile.}.
For the case of coh.dir, the exact result nicely agrees with EPA one Eq. (\ref{fgamma.coh.})
in the whole $p_{T}$ region. However, the result of Eq. (\ref{DZ.fgamma.coh.}) is much larger than
the results of exact treatment and Eq. (\ref{fgamma.coh.}), since Eq. (\ref{DZ.fgamma.coh.})
is obtained by neglecting the $m_{p}^{2}$ term in Eq. (\ref{fgamma.coh.}), and setting
$Q^{2}_{\textrm{max}}=\infty$ which will cause obvious errors. It can be seen that, the contribution
of res.pho is about two orders of magnitude larger than the dir.pho, thus the contribution of large
$p_{T}$ dileptons produced by photoproduction processes is mainly from res.pho. The case of coh.res
is similar to coh.dir, the errors from Eq. (\ref{DZ.fgamma.coh.}) is also prominent. Therefore, the
choice of $Q^{2}_{\textrm{max}}$ is crucial to the accuracy of EPA. Eq. (\ref{fgamma.coh.}) with
$Q^{2}_{\textrm{max}}=4p_{T}^{2}$ is a good choice for calculating coh.pho. However, choosing $Q^{2}\sim \infty$
will cause the large fictitious contribution from the large $Q^{2}$ domain (it can be found in the figures
which show the $Q^{2}$ dependence behaviour), which agree with the statements in Ref. \cite{Phys.Rep._15_181}.
For the practical use of EPA, except considering the kinematically allowed $Q^{2}$-change region,
one should also elucidate whether there is a dynamical cut off $\Lambda^{2}_{\gamma}$, and estimate it.
However, the definite values of the $\Lambda_{\gamma}^{2}$ for different processes are essentially
different, and still need further studies.

For the case of incoh.dir, the EPA results are greater than the exact one in
the whole $p_{T}$ region, but the difference between the EPA result Eq. (\ref{fgamma.incoh.noWF})
and the exact one is much more evident, since the $Q_{\textrm{max}}^{2}$ is set be $\hat{s}/4$ in
Eq. (\ref{fgamma.incoh.noWF}), which include the errors from the large $Q^{2}$ domain. The incoh.res
is similar to incoh.dir, the differences between the exact result and the EPA ones are prominent
in the whole $p_{T}$ region. Thus, the EPA approach can not be used in incoh.pho and the effects of
the inapplicability of EPA for incoh.pho are significant. In Fig. \ref{7PT.dile.} (c) and \ref{14PT.dile.} (c),
the comparisons between the photoproduction processes and the one of initial partons hard scattering are
presented. It can be seen that the corrections from the exact results of photoproduction processes to
had.scat are non-negligible, especially in the large $p_{T}$ domain. However, the differences between
the EPA results and the exact one are evident. The EPA results will give the large fictitious correction
to the production of dileptons and photons, especially for the EPA result of Eq. (\ref{DZ.fgamma.coh.})
and Eq. (\ref{fgamma.incoh.noWF}). Thus the statements are not accurate in Ref. \cite{Phys.Rev.C._84_044906, Phys.Rev.C._91_044908},
in which the incoh.pho of dileptons and photons was calculated by using the EPA approach Eq. (\ref{fgamma.incoh.noWF}) with
$Q^{2}_{\textrm{max}}=\hat{s}/4$ and $Q^{2}_{\textrm{min}}=1\ \textrm{GeV}^{2}$, and the coh.pho was
calculated by using the photon flux function with the DZ form Eq. (\ref{DZ.fgamma.coh.}).

Fig. \ref{Q2.pho.dir.} and \ref{Q2.pho.res.} present the $Q^{2}$ distribution of real photons produced by
photoproduction processes in p-p collision at LHC energies. It is shown that the differences among the exact
result and EPA ones are more obvious in the large $Q^{2}$ region. The $p_{T}$ distribution of real photons
produced by photoproduction processes in p-p collision at LHC energies can be found in Fig. \ref{7PT.pho.}
and Fig. \ref{14PT.pho.}. The results are similar to the case of dileptons in Fig. \ref{7PT.dile.} and \ref{14PT.dile.},
but the inapplicability of EPA for incoh.pho and the errors from the option of $Q^{2}_{\textrm{max}}\sim\infty$ and $Q^{2}_{\textrm{max}}=\hat{s}/4$
are more obvious. We also compare our results of real photons to Ref. \cite{Phys.Rev.C._84_044906, Phys.Rev.C._91_044908},
the inaccuracy of EPA is more evident. Therefore, the EPA can be used for coh.pho with the suitable choice of $Q^{2}_{\textrm{max}}$.
And for incoh.pho, EPA is not an effective treatment, since it dominates the large $Q^{2}$ region
where the errors are obvious. Thus, the exact treatment is needed to deal accurately with the
photoproduction of dileptons and photons.

\section{SUMMARY AND CONCLUSIONS}
We have investigated the production of large $p_{T}$ dileptons and photons induced by photoproduction processes
in p-p collisions at LHC energies, and presented the distributions of $Q^{2}$ and $p_{T}$. The exact treatment,
which returns to the EPA approach in the limit $Q^{2}\rightarrow 0$, is developed for calculations. The coherent
and incoherent photon emission processes are considered simultaneously and the Martin-Ryskin method is used for
avoiding the double counting. The comparisons between the exact results and EPA ones are presented for discussing
the applicability range of EPA and its accuracy. The numerical results indicate that, the EPA approach is only
a good approximation in the small $Q^{2}$ region and can be used for coh.pho with the suitable option of $Q^{2}_{\textrm{max}}$.
For incoh.pho, EPA is not an effective approximation, since the incoh.pho dominate the large $Q^{2}$
region where the errors are obvious. The photon flux function Eq. (\ref{DZ.fgamma.coh.}) developed by Drees and
Zeppenfeld and Eq. (\ref{fgamma.incoh.noWF}) with $Q^{2}_{\textrm{max}}=\hat{s}/4$ are widely used in the literatures
\cite{Nucl.Phys.B._904_386, Phys.Rev.C._92_054907, Chin.Phys.Lett._29_081301, Nucl.Phys.A._865_76, Phys.Rev.C._84_044906, Phys.Rev.C._91_044908},
and the the imprecise statements were given. Therefore, EPA can not provide accurate enough results for the
photoproduction of large-$p_{T}$ dileptons and photons in p-p collision, and the exact treatment should be considered.

\section*{ACKNOWLEDGMENTS}
We thank Yong-Ping Fu for useful communications. This work is supported in part by the National Natural Science
Foundation of China (Grant Nos. 11747086, 11465021, and 61465015), and by the Young backbone teacher
training program of Yunnan university. Z. M. is supported by Yunnan University's Research Innovation Fund for
Graduate Students (Grant No. YDY17108).

\end{document}